\pdfoutput=1
\documentclass{aa}
\titlerunning{Molecular cloud in 3C 84}
\usepackage{url}
\usepackage[utf8]{inputenc}
\DeclareUnicodeCharacter{2212}{\textminus}
\usepackage{CJKutf8}
\usepackage{xcolor}
\usepackage[normalem]{ulem}
\usepackage[switch]{lineno}
\usepackage{graphicx}
\usepackage{subfigure}
\usepackage{tikz}
\usetikzlibrary{arrows.meta}
\usepackage{txfonts}
\usepackage{natbib}
\usepackage{url}
\usepackage{hyperref}
\hypersetup
{
    colorlinks= true,
    linkcolor = blue,
    filecolor = blue,      
    urlcolor  = blue,
    citecolor = blue
}

\bibpunct{(}{)}{;}{a}{}{,} 
\begin{document}

   \title{Spatially resolved molecular gas conditions in the circumnuclear disc of 3C 84}
   \author{Bin Jia(\begin{CJK*}{UTF8}{gbsn}贾彬\end{CJK*})\inst{\ref{STRW}}\fnmsep\thanks{E-mail: \url{jia@strw.leidenuniv.nl}}, 
   Tom Oosterloo\inst{\ref{ASTRON},\ref{Kapteyn}},
   Serena Viti\inst{\ref{STRW},\ref{TRA},\ref{UCL}},
   Raffaella Morganti\inst{\ref{ASTRON},\ref{Kapteyn}}
          }

   \institute{
Leiden Observatory, Leiden University, P.O. Box 9513, 2300 RA Leiden, The Netherlands\label{STRW}
\and ASTRON, the Netherlands Institute for Radio Astronomy, Oude Hoogeveensedijk 4, 7991 PD, Dwingeloo, The Netherlands\label{ASTRON}
\and Kapteyn Astronomical Institute, University of Groningen, Postbus 800,
9700 AV Groningen, The Netherlands\label{Kapteyn}
\and Transdisciplinary Research Area (TRA) ‘Matter’/Argelander-Institut für Astronomie, University of Bonn, Bonn, Germany\label{TRA}
\and Department of Physics and Astronomy, University College London, Gower Street, London, UK\label{UCL}
}

    \abstract
    {The brightest cluster galaxy NGC~1275, at the centre of the Perseus cluster, hosts the radio-loud AGN 3C~84 and a circumnuclear disc (CND) of cold molecular gas. Large-scale molecular filaments traced by CO emission converge toward the nucleus and accrete onto this disc providing a direct observational link between gas cooling from the intracluster medium and the fuelling of the central engine. The physical and chemical conditions of the molecular gas within the CND, and how they are shaped by this accretion process, remain poorly constrained.}
    {We present a spatially resolved analysis of the molecular gas conditions across the CND aiming to determine the radial and azimuthal distributions of gas density, kinetic temperature, and cosmic ray ionisation rate, and to identify the processes that govern the observed molecular line emission.}
    {We analyse ALMA observations of CO(2--1), HCN(3--2), and HCO$^{+}$(3--2) at a resolution of 72\,pc. The data are partitioned into beam-sized hexagonal regions and modelled using a Bayesian inference framework that couples a neural network emulator of time-dependent chemistry (UCLCHEM) with non-LTE radiative transfer (SpectralRadex). For each region, we derive posterior distributions of the gas density, kinetic temperature, and cosmic ray ionisation rate. Bimodal posteriors are identified through Gaussian mixture modelling and validated with posterior predictive checks.}
    {All three line ratios peak in the inner disc and decline with radius. The Bayesian analysis shows radial gradients in gas density ($\log_{10}\,n(\mathrm{H_2}) \approx 6.3$ to $\sim 5$), kinetic temperature ($\sim200$\,K to $\sim160$\,K), and cosmic ray ionisation rate ($\log_{10}(\zeta/\zeta_0) \approx 4.9$ to $\sim 3$). Despite the powerful radio AGN, the observed HCN(3--2)/HCO$^{+}$(3--2) ratio remains $\lesssim 1$ across the disc; our modelling indicates that this is caused by optical depth saturation of HCN(3--2) ($\tau \sim 1$--3), which suppresses the intensity ratio even when the underlying HCN abundance exceeds that of HCO$^{+}$ by a factor of three or more. Azimuthally resolved profiles suggest a localised HCO$^{+}$/CO enhancement at the western disc boundary, coinciding with the filament--disc accretion interface.}
    {The HCN/HCO$^{+}$ intensity ratio cannot serve as a direct abundance diagnostic without accounting for optical depth, and a low observed ratio does not necessarily imply weak AGN influence on the chemistry. The western HCO$^{+}$/CO enhancement is consistent with shock processing driven by velocity shear between the infalling filaments and the rotating disc. These results indicate that the CND is shaped by accretion from the surrounding filamentary environment.}
    \keywords{galaxies: active -- galaxies: individual: 3C 84 -- galaxies: ISM -- ISM: molecules}
    \maketitle
%

\section{Introduction}

The Perseus cluster (Abell~426) is the brightest X-ray galaxy cluster in the sky and one of the nearest cool-core clusters, making it one of the most extensively studied systems in extragalactic astrophysics. At its centre, the brightest cluster galaxy NGC~1275 ($z = 0.01756$, $1\arcsec \approx 360$\,pc; \citealt{Huchra1999}) hosts the radio-loud AGN 3C~84, hereafter also used to refer to the host galaxy system. The radio source ejects jets of relativistic plasma to the north and south, which have been resolved on pc scales with very long baseline interferometry \citep[e.g.][]{Silver1998,Giovannini2018,Savolainen2023} and extend to kpc scales \citep[e.g.][]{Pedlar1990,Morganti2023}, inflating cavities and driving sound waves in the hot intracluster medium (ICM) that deposit mechanical energy offsetting radiative cooling \citep{Boehringer1993, Fabian2003, Fabian2006, Fabian2011}. Without this energy input the short cooling time of the ICM within the central tens of kiloparsec would lead to a massive cooling flow; yet, X-ray observations show that only a fraction of the expected cooling actually occurs \citep{Allen2001, Peterson2003}. Despite this regulation, significant reservoirs of cold gas survive in the cluster core, making Perseus a key laboratory for studying the cycle of AGN feedback, gas cooling, and black hole fuelling.

Observations over the past three decades have revealed that the central regions of 3C~84 harbour large quantities of cold molecular gas. Early detections of CO rotational emission toward the nucleus \citep{Lazareff1989, Mirabel1989, Reuter1993, Bridges1998} established the presence of a molecular reservoir of the order of $10^{10}$\,M$_\odot$ \citep{Salome2006}. Subsequent mapping with the IRAM 30\,m telescope showed that CO emission extends well beyond the galaxy centre, tracing a network of filaments out to projected distances of at least 50\,kpc, closely associated with the H$\alpha$-emitting nebula \citep{Conselice2001, Salome2006, Salome2008a, Salome2011}. Part of the molecular gas is found at the rims of the X-ray cavities excavated by the radio lobes, where the compressed ICM is thought to cool more efficiently \citep{Salome2006}. Warm molecular hydrogen detected in the near-infrared further traces the filamentary structures and shares the same kinematics as the cold CO gas \citep{Hatch2005, Wilman2005, Lim2012}. These spatial and kinematic correlations between the radio lobes, X-ray cavities, optical filaments, and molecular gas point to a close physical link between AGN activity and the formation or redistribution of cold gas in the cluster core.

On smaller scales, the question of how this filamentary gas connects to the central engine has been a long-standing problem. Using the Submillimeter Array (SMA), \citet{Lim2008} mapped CO(2--1) emission on kiloparsec scales and found filamentary structures aligned in the east--west direction with kinematics, suggestive of radial infall, although no clear signature of rotation was detected. At the much higher spatial resolution of the Atacama Large Millimeter/submillimeter Array (ALMA), \citet{Nagai2019} resolved a circumnuclear disc (CND) of cold molecular gas with a diameter of approximately 100\,pc, detected in CO(2--1), HCN(3--2), and HCO$^+$(3--2). The inner part of this disc exhibits fast Keplerian rotation with a position angle of $\sim 68\degr$, and its rotation axis is roughly aligned with the radio jet, suggesting a physical connection between the cold gas disc and the innermost accretion flow responsible for jet launching. This was the first evidence for a massive cold gas disc on this spatial scale in a bright cluster galaxy (BCG). More recently, \citet{Oosterloo2024} reprocessed the same ALMA data and recovered the extended emission that had been suppressed by calibration artifacts in the original reduction. The improved images revealed several kiloparsec-scale filaments converging toward the nucleus and showed, through their position--velocity structure, that these filaments are indeed accreting onto the CND. This result directly demonstrates that cold gas cooled from the ICM can flow back to the central regions and feed the supermassive black hole (SMBH), thereby closing the feedback--feeding loop.

The CND itself is a complex, multi-phase structure. Its inner fast-rotating region is embedded in a somewhat larger disc-like structure whose kinematics are less regular, reflecting the ongoing accretion of material from the surrounding filaments \citep{Scharwachter2013, Oosterloo2024}. In the warm molecular phase, \citet{Scharwachter2013} detected the CND in ro-vibrational H$_2$ and [Fe\,\textsc{ii}] emission and interpreted the observed velocity dispersion as indicative of a collisionally excited, turbulent accretion disc. Neutral atomic hydrogen has also been detected in the CND through broad H\,\textsc{i} absorption against the radio continuum at arcsecond resolution, with a line width and central velocity consistent with those of the molecular disc \citep{Morganti2023}. The background continuum providing the absorption is attributed to non-thermal synchrotron emission from star formation activity in the disc, whose presence has been independently confirmed through VLBI detections of diffuse synchrotron emission on scales of several tens of parsec \citep{Silver1998, Nagai2021}. At the same time, the AGN itself is classified as a low-excitation radio galaxy \citep{Buttiglione2009, Buttiglione2010} with an Eddington ratio of only $\sim 10^{-4}$ \citep{Punsly2018}, indicating that the radiative output of the accretion flow, and in particular the X-ray irradiation of the circumnuclear gas, is substantially weaker than in classical Seyfert nuclei. This combination of active star formation, filamentary accretion, and relatively weak radiative AGN output defines a distinctive environment in which the physical and chemical conditions of the molecular gas are shaped by multiple processes.

While CO traces the bulk molecular reservoir, species with higher critical densities such as HCN and HCO$^+$ are more sensitive to density gradients in the local physical conditions. In particular, their relative abundances and line ratios respond to the gas density, kinetic temperature, and the prevailing ionisation and heating mechanisms. Theoretical studies by \citet{Bayet2010} identified CN, HCO$^+$, and C$_2$H as tracers of regions influenced by cosmic ray ionisation and mechanical energy dissipation in the Perseus cluster environment. The observational follow-up by \citet{Bayet2011} provided the first detections of HCO$^+$(3--2) and CN(2--1) toward both 3C~84 and a position in the eastern filament, and comparison with chemical models indicated that cosmic ray heating at rates at least two orders of magnitude above the Galactic value is required to reproduce the observed molecular abundances. Earlier, \citet{Salome2008b} reported the first detection of HCN(3--2) toward the nucleus. These single-pointing observations established that dense molecular gas is present and that the heating environment differs markedly from that of the Galactic interstellar medium, but they could not resolve the spatial distribution of the physical conditions within the CND.

By combining time-dependent chemistry, non-LTE radiative transfer, and Bayesian inference, we constrain the gas density, kinetic temperature, and cosmic ray ionisation rate on scales of $\sim$70 pc across the disc. We examine the radial and azimuthal behaviour of the molecular line ratios, assess the role of optical depth in the interpretation of HCN emission relative to that of HCO$^+$, and identify a localised enhancement of HCO$^+$ relative to CO  at the western disc boundary that traces the interface between infalling filaments and the rotating disc. The paper is organised as follows. Section~\ref{sec:Observations} describes the observations and data reduction. Section~\ref{section:BayesianandNN} presents the modelling framework. Section~\ref{section:Results} presents the results, and Section~\ref{section:Discussion} discusses their implications. Section~\ref{section:conclusion} summarises our conclusions.


\section{ALMA observations and data reduction}
\label{sec:Observations}

\label{sec:ALMAObservations}

This paper is based on  re-processed archival CO(2--1), HCN(3--2) and HCO$^+$(3--2) observations of 3C 84 that were performed for ALMA project 2017.0.01257.S and of which the original data were published by \citet{Nagai2019}.  These authors used the data products produced by the standard ALMA pipeline of which the quality, as also discussed by them, is very poor because the line channels are dominated by frequency-dependent continuum residuals  due to  poor bandpass calibration. This poor bandpass calibration is due to the fact that, at the observed frequencies of 220--260 GHz,   3C 84 is one of the strongest compact sources in the sky ($\sim$7 Jy at the time of observations) and there was no bandpass calibrator available that was bright enough to provide a bandpass calibration of sufficient signal-to-noise when applied to the 3C 84 data. As a result, all channels of the ALMA pipeline data cubes are dynamic range limited due to residual continuum  errors covering the entire channels.

Profiting from the fact that the continuum emission of 3C 84 is much stronger than the CO(2--1) line emission, and that the continuum emission of 3C 84 at the relevant frequencies is a point source while the line emission is  extended, \citet{Oosterloo2024} were able to devise a spectral self-calibration technique  to improve the bandpass calibration of the CO(2--1) observations such  that  the frequency-dependent continuum errors are eliminated and that the data cubes are instead noise limited. The HCN(3--2) and HCO$^+$(3--2) observations of project 2017.0.01257.S suffer from the same calibration problem and  therefore we applied the same spectral self calibration method as used by \citet{Oosterloo2024} on the CO(2--1) data to the HCN(3--2) and HCO$^+$(3--2)  observations. As a result, the HCN(3--2) and HCO$^+$(3--2)  data cubes we present here are of much better quality than those produced by the ALMA pipeline and are noise limited. With the spectral response of all three cubes well calibrated, the continuum of 3C~84 varies smoothly across the line-free channels and is removed with a linear fit that leaves no residual or over-subtraction in any line. An impression of the quality of the  data cubes is given in Appendix \ref{sec:appData}. 

The HCN(3--2) and HCO$^+$(3--2) data cubes were made using exactly the same steps as used for the CO(2--1) data cube as published by \citet{Oosterloo2024}. As a final step, the data cubes of all three emission lines were smoothed to a common spatial resolution of 0\farcs2 (corresponding to 72 pc) and re-gridded to the same spatial and velocity grids. The spectral resolution of these final cubes is $\sim$12 km s$^{-1}$ and the noise levels are 0.36, 0.30, and 0.41 mJy beam$^{-1}$ for CO(2--1), HCN(3--2), and HCO$^{+}$(3--2), respectively.

\section{Bayesian and neural network workflow}
\label{section:BayesianandNN}
   
To characterize the molecular gas in the circumnuclear disc of 3C\,84 and to quantify how its physical conditions vary across the system, we developed a modeling framework that combines time dependent chemistry, radiative transfer, and Bayesian inference. The chemical abundances are computed with \texttt{UCLCHEM} \citep{holdship2017}\footnote{\url{https://github.com/uclchem/UCLCHEM}}, a time dependent gas grain chemical code that uses user defined chemical networks and physical modules to simulate a wide range of interstellar environments. Radiative transfer calculations are performed with \texttt{SpectralRadex} \citep{Holdship2021}\footnote{\url{https://spectralradex.readthedocs.io}}, which is based on \texttt{RADEX} \citep{radex} and predicts molecular line intensities from given column densities and physical conditions using collisional rate coefficients from the LAMBDA database \citep{Lambda2005}. 

\begin{figure*}
    \centering
    \begin{tikzpicture}[
        box/.style={rectangle, rounded corners, draw=black, minimum width=4cm, minimum height=1.5cm, text centered, text width=4cm, align=center},
        arrow/.style={->, >=latex, thick},
        node distance=3cm
    ]
    
    \node[box, fill=blue!15] (A) at (0,0) {Parameter Selection \\ \small{Choose physical parameters }};
    
    \node[box, fill=red!15] (B) at (7,2) {Abundance Prediction \\ \small{Neural network estimates molecular abundances}};
    
    \node[box, fill=blue!15] (C) at (7,-2) {Line Intensity Calculation \\ \small{SpectralRadex computes model intensities}};
    
    \node[box, fill=blue!15] (D) at (13,0) {Likelihood Evaluation \\ \small{Compare model with observations}};
    
    \draw[arrow] (A) -- node[above, sloped] {Pass $T_{k}$,$n$,$\zeta$} (B);
    \draw[arrow] (A) |- node[above, near end] {Pass $T_{k}$,$n$,$N_{H_{2}}$} (C);
    \draw[arrow] (B) -- node[right] {Pass $X_{mol}$} (C);
    \draw[arrow] (C) -- node[above, sloped] {Line Intensity} (D);
    
    \draw[arrow] (D.north) -- ++(0,2) -| node[above, near end] {Update parameters} (A.north);
    
    \end{tikzpicture}
    \caption{Workflow of our Bayesian analysis. The neural network component (red) accelerates the analysis by replacing computationally intensive chemical modeling. Each box shows a key step in the process, with arrows indicating data flow between steps. The feedback loop allows continuous refinement of parameter estimates until convergence is achieved.}
    \label{fig:workflow}
    \end{figure*}
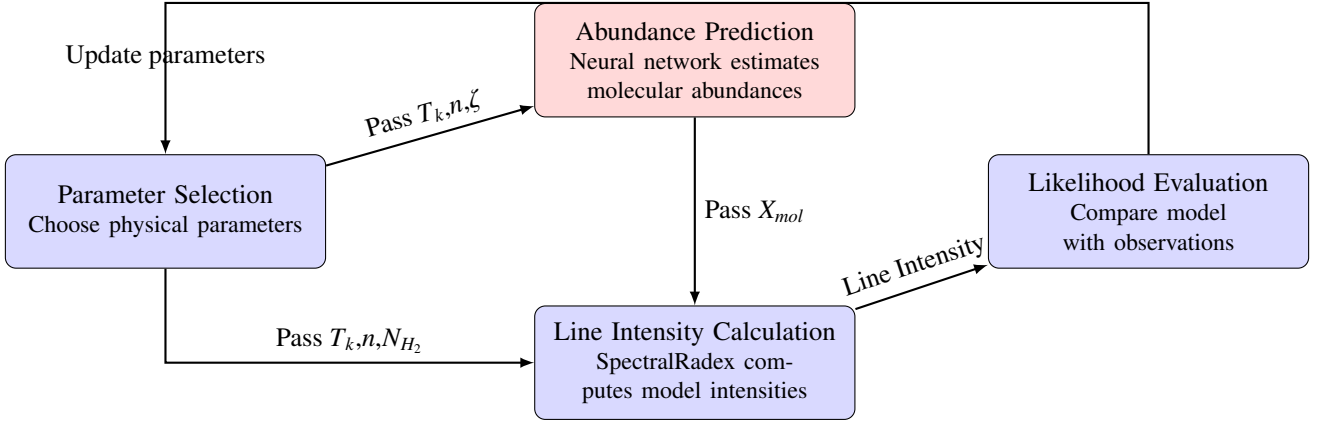

Directly coupling chemical simulations with radiative transfer within a Bayesian framework is computationally expensive, as a large number of model evaluations is required to sample the parameter space. To address this, we employ a pre-trained neural network emulator, trained on a grid of chemical models, to approximate the mapping between physical conditions and molecular abundances. For each set of physical parameters, the emulator predicts the fractional abundances of CO, HCN, and HCO$^{+}$ as a function of gas density, kinetic temperature, and cosmic-ray ionisation rate. These abundances are multiplied by the total molecular hydrogen column density $N_{\rm H_2}$ to obtain the corresponding molecular column densities. The resulting column densities, together with the gas density and kinetic temperature, are then supplied to \texttt{SpectralRadex} to compute model integrated intensities for the observed transitions. The comparison between model predictions and observations is carried out within a Bayesian framework using \texttt{emcee}, which is used to explore the posterior distributions of the physical parameters. Each likelihood evaluation therefore follows a forward sequence in which a trial set of physical parameters is first mapped to molecular abundances through the emulator, then converted to column densities, and finally translated into observable line intensities via non-LTE radiative transfer before being compared with the data. The overall workflow is illustrated in Fig.~\ref{fig:workflow}, and since the neural network architecture closely follows that presented in Jia et al.\ (submitted), we do not repeat the details of the training procedure here.

\begin{table*}
\caption{Parameter range and posterior distribution in $\chi^{2}$-fitting and the Bayesain analysis}
\label{table:priors}
\centering
\begin{tabular}{lcccc}
\hline\hline
Parameter & Description & Range & Distribution type & Number of grid points  \\
\hline
$T_{\mathrm{K}}$ & Temperature (K) & $50$ to $600$ & Uniform & $26$ \\
$n_{\mathrm{H}_2}$ & Volume density (cm$^{-3}$) & $10^{3}$ to $10^{7}$ & Logarithmic & $23$ \\
$\zeta$ & Cosmic ray ionisation rate ($\zeta_0$)\tablefootmark{a} & $10^{2}$ to $10^{6}$ & Logarithmic & $31$ \\
$N_{\mathrm{H}_2}$ & H$_2$ column density (cm$^{-2}$) & $10^{22}$ to $10^{25}$  & Logarithmic & $26$ \\
\hline
\end{tabular}
\tablefoot{
\tablefootmark{a} $\zeta_0 = 1.36 \times 10^{-17}$ s$^{-1}$.
The ranges in this table define the priors used in the Bayesian analysis. The UCLCHEM grid spans at least these ranges in $n_{\rm H_2}$, $T_{\rm k}$, and $\zeta$ so that the emulator can be evaluated across the full prior volume. The H$_2$ column density is not varied in the UCLCHEM models and is treated only in the radiative transfer step.
}
\end{table*}

\subsection{Bayesian analysis with intensity-based likelihood}
\label{subsection:BayesAnalysis}

The integrated intensity maps of HCN(3--2), HCO$^{+}$(3--2), and CO(2--1) are constructed at a common spatial resolution of 72 pc.  The moment--0 maps are obtained by applying a $3\sigma$ clipping threshold and integrating only emission above this level. The integration is carried out over $|v - v_{\rm sys}| \leq 300\ \mathrm{km\ s^{-1}}$ to include both the bulk rotating gas and the infalling component associated with the filaments \citep{Oosterloo2024}.

The uncertainty on the integrated intensity, $\sigma_{\rm mom0}$, is estimated from the rms noise per channel, $\sigma_{\rm chan}$, the number of contributing channels, $N_{\rm chan}$, and the channel width, $\Delta V$, such that $\sigma_{\rm mom0} = \sigma_{\rm chan}\,\sqrt{N_{\rm chan}}\,\Delta V$. To account for systematic effects, a calibration uncertainty of 15\% is further included and added in quadrature to the statistical uncertainty \citep{Francis2020}.  Regions in which any of the three transitions is not detected, or where the measured intensity falls below $3\sigma_{\rm mom0}$, are excluded from the parameter inference, so the analysis uses only hexagons with all three lines detected above $3\sigma$.

To enable spatially resolved analysis, the moment--0 maps are partitioned into adjoining hexagonal regions using the HExagonal Region Averager (HERA)\footnote{https://github.com/ebehrens97/HERA} code \citep{erica_behrens_2024_13839853}. Each hexagon has a characteristic size of 72\,pc, corresponding to the $\sim0\farcs2$ beam, such that adjacent regions are approximately independent. The integrated intensities of all available transitions of CO, HCN, and HCO$^{+}$ are extracted within each hexagon. We show in Appendix~\ref{app:kinematics} that the three transitions are co-moving within each hexagon, so their integrated intensities can be treated as constraints on a single emitting region.

These measurements are then used as input for the Bayesian analysis, where the observed integrated intensities are compared with the predictions from the neural network emulator to infer the physical conditions of the dense molecular gas in each region of the circumnuclear disc.

The likelihood is evaluated directly in intensity space. For a detected transition with observed integrated intensity $I_{\rm obs}$, uncertainty $\sigma_{\rm obs}$, and model prediction $I_{\rm mod}$, we assume independent Gaussian errors and write the contribution to the log--likelihood as
\begin{equation}
\ln \mathcal{L}_{\rm det}
=
-\frac{(I_{\rm obs}-I_{\rm mod})^{2}}{2\sigma_{\rm obs}^{2}}
-\ln\!\left(\sqrt{2\pi}\,\sigma_{\rm obs}\right).
\end{equation}

The total likelihood is the product of the contributions from all detected transitions.

The inference is carried out in two stages in order to reduce the dimensionality of the Bayesian parameter space. Because only three transitions are available, constraining four free parameters within a single Bayesian inference would lead to poorly determined posteriors. We therefore first fix the column density through a separate step. In the first stage, we apply a $\chi^{2}$ minimization using the parameter grid listed in Table~\ref{table:priors}, allowing all four parameters, including $\log_{10} N_{\rm H_{2}}$, to vary freely. This step yields the best--fitting value of $\log_{10} N_{\rm H_{2}}$, which is adopted as a reference column density for each region. In the second stage, we fix $\log_{10} N_{\rm H_{2}}$ to this reference value and perform the Bayesian inference with ${\boldsymbol{\theta}}=(\log_{10} n({\rm H_{2}}),\,T_{\rm gas},\,\log_{10}(\zeta/\zeta_{0}))$ as the three remaining free parameters. To assess the robustness of this assumption, we further explore a range of $\log_{10} N_{\rm H_{2}}$ values around the reference point. Specifically, we sample values within a narrow interval, spaced by 0.2 dex, and repeat the Bayesian inference for each case. This procedure allows us to evaluate the sensitivity of the inferred parameters to the adopted column density. The results of this analysis are presented in Appendix~\ref{sec:appx_Nh2check}.

The posterior probability distribution is written as
\begin{equation}
p({\boldsymbol{\theta}} \mid {\bf I})
\propto
\mathcal{L}({\bf I} \mid {\boldsymbol{\theta}})\,p({\boldsymbol{\theta}}),
\end{equation}
\noindent where ${\bf I}$ denotes the observed integrated line intensities and $p({\boldsymbol{\theta}})$ represents the prior distributions of the free parameters.
The likelihood function $\mathcal{L}({\bf I} \mid {\boldsymbol{\theta}})$ is evaluated in intensity space as described above, and the column density is not treated as a free parameter at this stage.

The resulting posterior distributions provide constraints on the physical conditions of the molecular gas. For each 72\,pc hexagonal region, we derive marginal posterior distributions for $\log_{10} n({\rm H_{2}})$, $T_{\rm gas}$, and $\log_{10}(\zeta/\zeta_{0})$ (Fig.~\ref{fig:corner_examples}), and we quote uncertainties using the central 68\% credible intervals.

\subsection{Bimodality identification and posterior predictive checks}
\label{subsec:bimodal_ppc}

\begin{figure*}
    \centering
    \includegraphics[width=\textwidth]{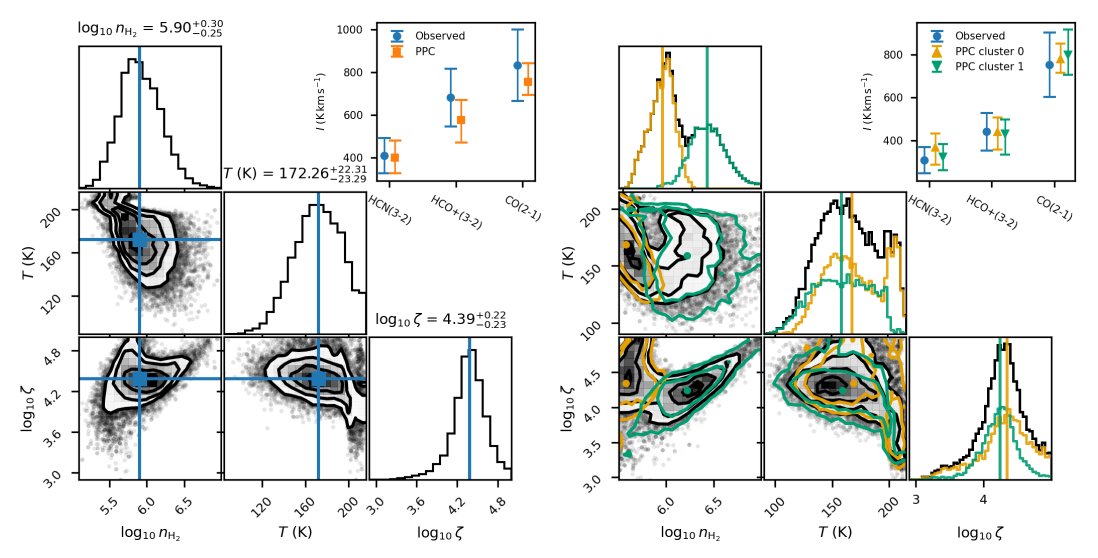}
    \caption{Examples of posterior distributions obtained from the Bayesian analysis. 
    Left: corner plot for region 14, highlighted in yellow in Fig.~\ref{fig:physical_parameter_maps}, showing a  well constrained posterior component, with the top right panel presenting the PPC comparing the observed and predicted line intensities. 
    Right: corner plot for region 16, highlighted in magenta in Fig.~\ref{fig:physical_parameter_maps}, where the posterior distribution separates into two components. The two clusters are highlighted in green and yellow, and the PPC panel shows the predicted intensities drawn from each component.}
    \label{fig:corner_examples}
\end{figure*}

For four of the mapped regions, the posterior distributions of the physical parameters exhibit clear bimodal structures, as illustrated in the right panel of Fig.~\ref{fig:corner_examples}, where the density posterior displays two distinct peaks. To characterise these cases, and to assess whether both solutions remain consistent with the observations, we perform an additional analysis consisting of two steps, namely the identification of bimodal clusters in parameter space followed by posterior predictive checks (PPC) carried out separately for each cluster.

For regions with bimodality, we use the posterior samples obtained from the second stage of the Bayesian inference, which provide samples of the parameter vector
$\boldsymbol{\theta}=(\log_{10} n_{\mathrm{H_2}},\, T_{\rm gas},\, \log_{10} \zeta)$.
These samples are flattened and fitted with a two-component Gaussian mixture model (GMM). The model estimates the weights, means, and covariance matrices of two multivariate Gaussian components in the three-dimensional parameter space. The GMM therefore provides an empirical description of the two modes in the posterior distribution. For each component we record the mean physical parameters and the associated covariance matrix. Given the fitted mixture model, we compute, for each posterior sample, the posterior probability of belonging to each component. These probabilities quantify the degree to which a given sample is associated with one of the two clusters. They are subsequently used to draw predictive samples corresponding to each mode.

To evaluate whether each cluster provides a physically consistent solution, we perform PPC using the forward model adopted in the Bayesian inference. For each region and each GMM component, we draw a set of parameter samples from the posterior distribution using the component responsibilities as sampling weights. Each sampled parameter vector $(n_{\mathrm{H_2}}, T_{\mathrm{K}}, \zeta)$, together with the column density determined in the first stage of the analysis, is passed through the forward model to predict the line intensities for the observed transitions.
From these predictive realizations we compute the posterior predictive distribution for each transition and summarize it using the 16th, 50th, and 84th percentiles. The predicted distributions are then compared with the observed line intensities and their uncertainties. This procedure allows us to verify whether both modes of the posterior distribution are capable of reproducing the observed data. The PPC therefore serves as a consistency check for bimodal regions. If both clusters reproduce the observations within uncertainties, the bimodality reflects genuine degeneracies in the parameter space rather than numerical artifacts of the inference.

\section{Results}
\label{section:Results}
In this section we present the observational results and the physical parameters derived from the Bayesian analysis. Section~\ref{sec:lineimageratio} describes the integrated intensity maps and the spatial distribution of the molecular line ratios. Section~\ref{subsection:parametermaps} presents the gas density, kinetic temperature, and cosmic ray ionisation rate maps obtained from the \texttt{UCLCHEM}--\texttt{SpectralRadex} modelling framework, together with the predicted optical depths.

\subsection{Molecular line emission and ratios}
\label{sec:lineimageratio}

\begin{figure*}[htbp]
    \centering
    \includegraphics[width=\textwidth]{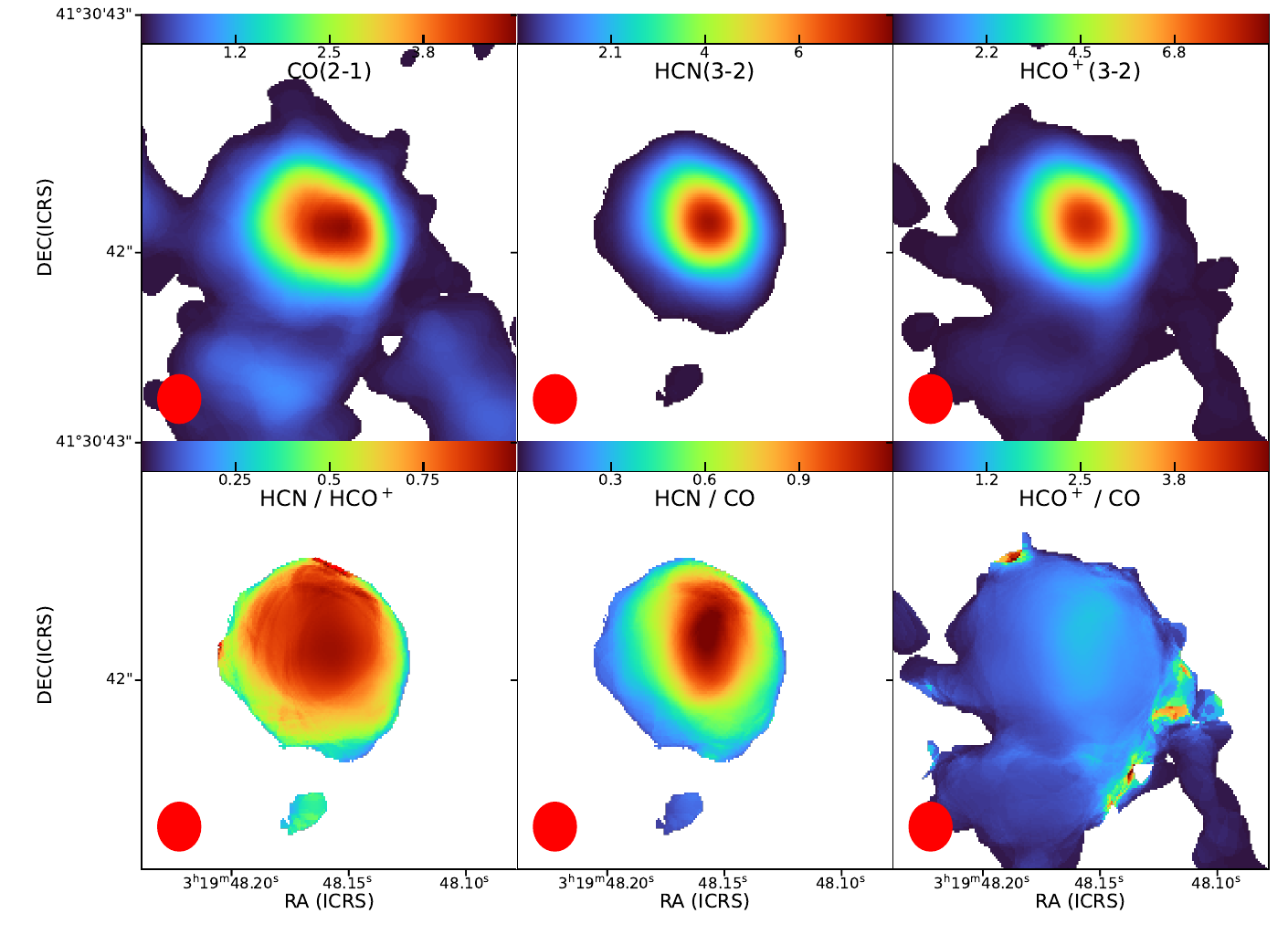}
    \caption{The top row shows the moment-0 maps of CO(2--1), HCN(3--2), and HCO$^+$(3--2) in Jy\,beam$^{-1}$\,km\,s$^{-1}$, integrated over $|v-v_{\rm sys}|\le300$\,km\,s$^{-1}$, while the bottom row shows the corresponding line-ratio maps HCN/HCO$^+$, HCN/CO, and HCO$^+$/CO. The ratios are computed after converting the integrated intensities from Jy\,beam$^{-1}$\,km\,s$^{-1}$ to K\,km\,s$^{-1}$. The conversions are 0.60, 0.45, and 0.44\,K per mJy\,beam$^{-1}$ for CO(2--1), HCN(3--2), and HCO$^+$(3--2) at their observed frequencies.The rms noise of the moment-0 maps is 31.6, 24.6, and 33.7\,mJy\,beam$^{-1}$\,km\,s$^{-1}$ for CO(2--1), HCN(3--2), and HCO$^+$(3--2), respectively. The synthesised beam is indicated in the lower-left corner of each panel.}
    \label{fig:3mom0_3ratio_72pc}
\end{figure*}

Figure~\ref{fig:3mom0_3ratio_72pc} (Top row) presents the integrated intensity maps of HCN(3--2), HCO$^{+}$(3--2), and CO(2--1). To ensure uniform spatial resolution across all directions, the data were smoothed to a circular beam of $0\farcs2 \times 0\farcs2$, corresponding to a physical scale of 72\,pc. All three transitions trace a compact molecular structure associated with the central region, with emission peaks broadly coincident in position. The CO(2--1) emission is the most extended component and outlines the full molecular reservoir in the circumnuclear disc and its environment (for more details on the CO(2--1), see \citealt{Oosterloo2024}). In contrast, the HCN(3--2) and HCO$^{+}$(3--2) emission are more centrally concentrated, indicating that these transitions preferentially trace denser gas within the same structure. The overall morphology and the kinematics of the dense gas tracers follow that of CO, but with a sharper radial decline in intensity. This spatial segregation between CO and the dense gas tracers is consistent with the findings of \citet{Nagai2019}, who first reported that CO traces both the disc and its immediate surroundings while HCN and HCO$^{+}$ are confined to the inner disc, based on the original reduction of the same ALMA dataset. The improved calibration applied in this work \citep{Oosterloo2024} recovers more extended CO emission but does not alter the central concentration of the dense gas tracers.

\begin{figure}[htbp!]
  \centering
  \includegraphics[width=\columnwidth]{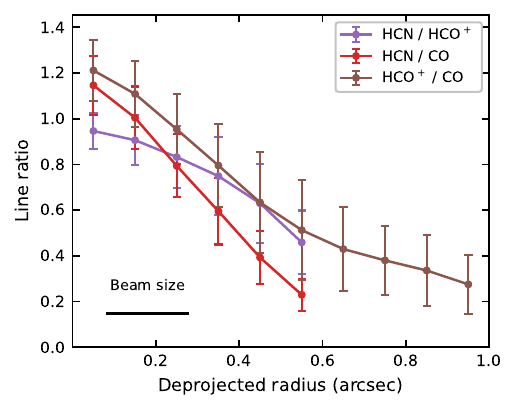}
  \caption{Azimuthally averaged radial profiles of the
    $\mathrm{HCN}/\mathrm{HCO}^{+}$, $\mathrm{HCN}/\mathrm{CO}$, and
    $\mathrm{HCO}^{+}/\mathrm{CO}$ intensity ratios as a function of
    deprojected radius. The disc geometry is defined by
    $\mathrm{PA} = 68^{\circ}$ and $i = 45^{\circ}$. The error bars
    indicate the $1\sigma$ scatter within each annulus.}
  \label{fig:ratio_radial_profile}
\end{figure}

Despite the overall similarity between the two dense gas tracers, the HCO$^{+}$(3--2) emission is systematically more extended than HCN(3--2), particularly toward the outer disc. This difference could be related to excitation conditions. At $T_{\mathrm{k}} = 100$~K, the critical density of HCO$^{+}$(3--2), $n_{\mathrm{crit}} \simeq 8.1 \times 10^{5}\,\mathrm{cm^{-3}}$, is significantly lower than that of HCN(3--2), $n_{\mathrm{crit}} \simeq 3.8 \times 10^{6}\,\mathrm{cm^{-3}}$ \citep{Shirley_2015}. As a result, HCO$^{+}$(3--2) can be efficiently excited over a wider range of gas densities, including regions where HCN(3--2) remains sub-thermally excited. 

Figure~\ref{fig:3mom0_3ratio_72pc} (Bottom row) presents the line ratio maps of $\mathrm{HCN}/\mathrm{HCO}^{+}$, $\mathrm{HCN}/\mathrm{CO}$, and $\mathrm{HCO}^{+}/\mathrm{CO}$ derived from the integrated intensity maps. Across most of the disc the $\mathrm{HCN}/\mathrm{HCO}^{+}$ ratio remains close to or below unity and exhibits a gradual decrease with increasing radius. Elevated $\mathrm{HCN}/\mathrm{HCO}^{+}$ intensity ratios, in particular values well above unity, are often reported toward AGN-dominated nuclei and have been attributed to enhanced HCN abundance or excitation under strong X-ray irradiation \citep[e.g.][]{Meijerink2007, Martin2015,Imanishi2020,Nakajima2023,Josh2025}. The absence of such an enhancement in 3C~84, despite its powerful radio AGN, suggests that additional factors shape the observed ratio. We return to this point in Sect.~\ref{subsec:opacity_hcn_hcop}.

To quantify the radial dependence, we deproject the disc assuming a position angle $\mathrm{PA} = 68^{\circ}$ \citep{Oosterloo2024} and an inclination $i = 45^{\circ}$ \citep{Scharwachter2013}, and compute azimuthally averaged intensity ratios in concentric annuli. Figure~\ref{fig:ratio_radial_profile} presents the resulting radial profiles of $\mathrm{HCN}/\mathrm{HCO}^{+}$, $\mathrm{HCN}/\mathrm{CO}$, and $\mathrm{HCO}^{+}/\mathrm{CO}$. All three ratios peak in the inner disc and decline steadily outward. The $\mathrm{HCO}^{+}/\mathrm{CO}$ ratio is the highest at all radii, decreasing from $\sim 1.2$ in the centre to $\sim 0.3$ at the disc periphery, while $\mathrm{HCN}/\mathrm{CO}$ follows a similar trend with consistently lower values. The $\mathrm{HCN}/\mathrm{HCO}^{+}$ ratio drops from $\sim 0.95$ near the centre to $\sim 0.4$ at a deprojected radius of $\sim 0\farcs6$, confirming that $\mathrm{HCO}^{+}$ increasingly dominates the dense-gas emission toward the outer disc.

These values are notably higher than the dense-gas-to-CO ratios measured in nearby star-forming galaxies, where spatially resolved $\mathrm{HCN}(1{-}0)/\mathrm{CO}(2{-}1)$ and $\mathrm{HCO}^{+}(1{-}0)/\mathrm{CO}(2{-}1)$ ratios typically remain below $\sim 0.07$ and $\sim 0.05$, respectively, even in the central regions \citep{GarciaRodriguez2023}. Among AGN hosts, the circumnuclear disc of NGC~1068 exhibits $\mathrm{HCN}(1{-}0)/\mathrm{CO}(1{-}0) \sim 0.4$--$1$, whereas starburst or composite systems generally show values below $\sim 0.2$--$0.3$ \citep{Krips2008, Sanchez2022}. The strong central concentration of both $\mathrm{HCN}/\mathrm{CO}$ and $\mathrm{HCO}^{+}/\mathrm{CO}$ in 3C~84 therefore indicates that the molecular medium becomes progressively dominated by dense and highly excited gas toward the inner disc, although the absolute values should be interpreted with caution because these ratios compare transitions with different critical densities.

These spatial trends in emission and line ratios suggest that the excitation conditions vary systematically across the disc and cannot be captured by a single set of physical parameters. The central peaking of all three ratios is qualitatively consistent with a radial decline in gas density and excitation toward the outer disc, while the persistently high HCO$^{+}$/CO values suggest that dense or strongly excited gas contributes significantly at all radii. In the next section, we quantify these trends by applying our Bayesian non-LTE chemical analysis to the observed line intensities across the mapped region, deriving spatially resolved distributions of gas density, kinetic temperature, and cosmic ray ionisation rate.

\subsection{Physical properties of the cloud}
\label{subsection:parametermaps}
\begin{figure*} 
\centering 
\includegraphics[width=0.85\textwidth]{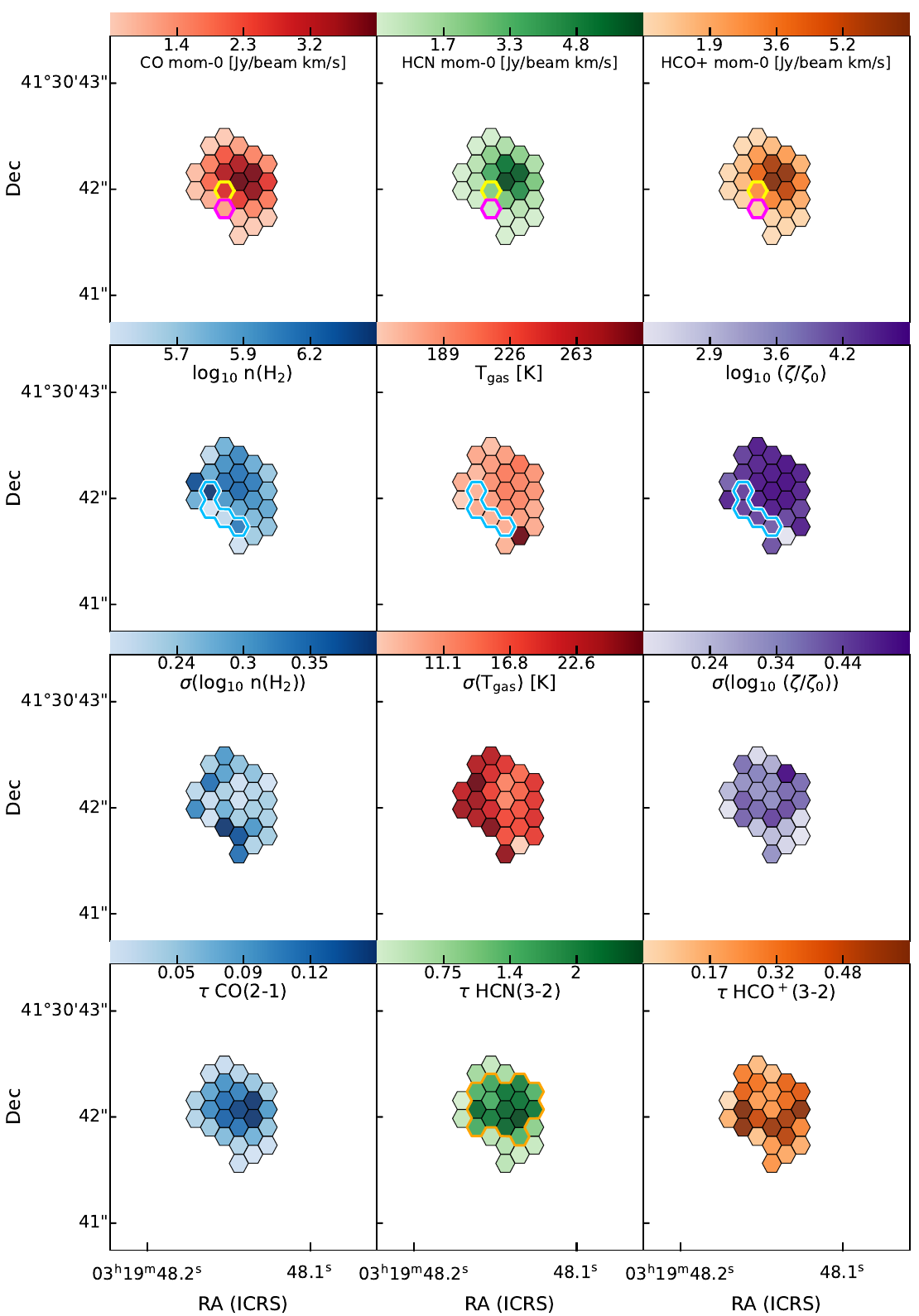} 
\caption{Hexagon-binned maps of molecular emission and derived gas properties in the central region of 3C\,84. Each hexagon corresponds to a beam-sized region used in the analysis. The top row shows the integrated intensity maps of CO(2--1), HCN(3--2), and HCO$^{+}$(3--2). The yellow and magenta outlined hexagons mark the two regions whose posterior distributions are shown in the left and right panels of Fig.~\ref{fig:corner_examples}, respectively. The second row presents the inferred physical parameters from the Bayesian analysis: $\log_{10} n(\mathrm{H_2})$, $T_{\rm gas}$, and $\log_{10}(\zeta/\zeta_0)$. Parameters are shown only where all three lines are detected above $3\sigma$. A single non-detection does not yield a reliable constraint. Cyan outlines mark the four bimodal regions. The third row shows the corresponding uncertainties. The bottom row shows the predicted optical depths for CO(2--1), HCN(3--2), and HCO$^{+}$(3--2). Hexagons outlined in orange mark regions where the predicted optical depth exceeds unity.}
\label{fig:physical_parameter_maps} 
\end{figure*}

We first assess the physical conditions of the molecular gas through Bayesian inference. Across the disc the posterior distributions are either unimodal or bimodal. The inference yields a single well-constrained mode in most regions, while four regions are bimodal (Sect.~\ref{subsec:bimodal_ppc}). Figure~\ref{fig:corner_examples} shows one region of each type, selected as clear examples of the two cases. The left panel shows the corner plot for a well constrained region (yellow hexagon in Fig.~\ref{fig:physical_parameter_maps}), where the marginalised posteriors yield $\log_{10} n({\rm H_{2}}) = 6.0^{+0.4}_{-0.3}$, a kinetic temperature of $T_{\rm gas} = 180^{+22}_{-23}$\,K, and a cosmic ray ionisation rate of $\log_{10}(\zeta/\zeta_{0}) = 4.6^{+0.2}_{-0.4}$. The top right panel of this example shows the PPC, where the observed line intensities are compared with those predicted from the posterior samples for CO(2--1), HCN(3--2), and HCO$^{+}$(3--2). In some regions, however, the posterior distribution is bimodal, as illustrated in the right panel of Fig.~\ref{fig:corner_examples} (magenta hexagon in Fig.~\ref{fig:physical_parameter_maps}). For these cases, we apply the Gaussian mixture decomposition and posterior predictive checks described in Sect.~\ref{subsec:bimodal_ppc} and mark the two solutions with cyan outlines in Fig.~\ref{fig:physical_parameter_maps}, colouring each hexagon by its dominant component.

Fig.~\ref{fig:physical_parameter_maps} presents a four-row summary of the observational constraints and the inferred physical conditions across the circumnuclear region. The first row shows the hexagonalized integrated intensity maps of CO(2--1), HCN(3--2), and HCO$^{+}$(3--2). The second row displays the corresponding best-fitting physical parameters derived from the Bayesian analysis, including $\log_{10} n({\rm H_{2}})$, $T_{\rm gas}$, and $\log_{10}(\zeta/\zeta_0)$. The third row shows the $1\sigma$ uncertainty on each parameter, taken as the half-width of the 68\% credible interval.The bottom row shows the optical depth $\tau$ predicted for each transition based on the best-fitting solutions.

The gas density exhibits a strong central concentration, with the inner disc characterised by a very dense environment with $\log_{10} n({\rm H_{2}}) > 6$. The highest density is found in the central region, reaching $\log_{10} n({\rm H_{2}}) = 6.3^{+0.3}_{-0.3}$. Moving outward from the nucleus, the disc still maintains relatively high densities, with typical values around $\log_{10} n({\rm H_{2}}) \sim 5$. This sustained dense environment is consistent with the density gradient proposed by \citet{Nagai2021} and is consistent with a scenario in which filamentary structures efficiently transport molecular material into the disc, leading to gas accumulation consistent with recent observational results \citep{Oosterloo2024}.

The spatial distribution of the gas temperature exhibits a radial profile similar to that of the density. We find the highest temperatures in the central region, reaching $T_{\rm gas} \sim 200^{+10}_{-19}$ K. Moving outward from the nucleus, the temperature gradually decreases to typical values of $T_{\rm gas} \sim 160^{+20}_{-30}$ K in the outer parts of the disc. The elevated temperatures in this inner region are likely driven by a combination of direct radiative heating from the active galactic nucleus and mechanical energy injected by the radio jet. As the influence of these central heating sources diminishes with distance, the outer disc is characterized by comparatively cooler molecular gas.

The cosmic ray ionisation rate is strongly enhanced throughout the disc and exhibits a clear radial gradient. In the central inner disc, we find $\log_{10}(\zeta/\zeta_{0}) > 4$, reaching a maximum of $\log_{10}(\zeta/\zeta_{0}) = 4.9^{+0.3}_{-0.2}$, corresponding to $\zeta \approx 10^{-12}$\,s$^{-1}$, among the highest values reported in any extragalactic molecular environment. The total ionisation power implied by this rate in the inner disc is $P_{\rm ion} \approx 10^{44}$\,erg\,s$^{-1}$\footnote{We estimate $P_{\rm ion} = \zeta \, E_{\rm ion} \, n_{\rm H} \, V$, adopting a mean energy deposition of $E_{\rm ion} \approx 30$\,eV per ionisation event in molecular gas \citep{Padovani2009}, $n_{\rm H} = 2\,n({\rm H}_2) \approx 2 \times 10^{6}$\,cm$^{-3}$, and a disc volume $V = \pi r^{2} h$ with $r = 70$\,pc and $h = 5$\,pc \citep{Nagai2021}.}, which amounts to roughly $1$--$10$\,\% of the jet kinetic power $L_{\rm jet} \sim 10^{45}$--$10^{46}$\,erg\,s$^{-1}$ \citep{Fabian2011, Kino2021}, indicating that the jet is energetically capable of sustaining the observed rate. By contrast, the circumnuclear star formation provides a cosmic ray luminosity of only $L_{\rm CR,SN} \sim 10^{41}$\,erg\,s$^{-1}$\footnote{The supernova-driven cosmic ray luminosity is estimated as $L_{\rm CR,SN} = \eta_{\rm CR} \, R_{\rm SN} \, E_{\rm SN}$, with a supernova rate $R_{\rm SN} \approx 0.03$\,yr$^{-1}$ \citep{Silver1998, Diehl2006}, $E_{\rm SN} = 10^{51}$\,erg \citep{Morlino2012}, and a canonical acceleration efficiency $\eta_{\rm CR} = 10$\,\%.}, falling short of $P_{\rm ion}$ by three orders of magnitude. Our spatially resolved gradient also extends the findings of \citet{Bayet2011}, who inferred $\zeta/\zeta_{0} \gtrsim 100$ from unresolved single-dish observations toward 3C~84; the outer disc values of $\log_{10}(\zeta/\zeta_{0}) \sim 3$ are consistent with that estimate once the beam dilution of JCMT is taken into account, while the central regions exceed it by two to three orders of magnitude. A notable exception to the radial decline is found on the western side of the disc, at distances of $\sim 200$--$300$\,pc, where $\log_{10}(\zeta/\zeta_{0}) \gtrsim 4$. At these distances neither the diluted jet contribution nor the supernova-driven cosmic ray budget can account for the enhancement, pointing to an additional local energy source; we discuss its likely origin in Sect.~\ref{subsection:HCO+/CO discussion}.

The radial behaviour described above is derived from a forward model in which the column density of each region is fixed in the first stage of the inference. The uncertainties given above are statistical only. In Appendix~\ref{sec:appx_Nh2check} we repeat the inference with $N_{\rm tot}$ shifted by $\pm0.2$ and $\pm0.4$ dex and keep only the cases that still pass the posterior predictive check (see Fig.~\ref{fig:galleryofNtotcheck}). The accepted variations define the systematic range of each parameter, shown together with the radial profiles in Fig.~\ref{fig:ntot_radial}. The origin of this systematic range differs between parameters. For the density, most of the scatter comes from the bimodal posteriors resolved by the GMM fit. For the temperature and the ionisation rate, it is driven by the column density itself, and the highest accepted values of both arise from shifted $N_{\rm tot}$ fits whose posteriors collapse into narrow spikes or pile up against the prior boundaries, so these extremes should be treated with caution. Despite this scatter, the radial decline of all three parameters is preserved across the full range of accepted $N_{\rm tot}$. The declining trends are therefore robust to the choice of column density, while the value found for any single hexagon is model dependent. It is the best estimate that a forward model can return from three observed lines, and not a direct measurement of the local conditions.

\section{Discussion}
\label{section:Discussion}

The results presented above reveal a complex picture of the molecular gas in the circumnuclear region of 3C~84. The circumnuclear disc does not appear as a dynamically isolated structure, but instead interacts with the surrounding molecular environment where large-scale filaments approach and connect to the disc, confirming the picture suggested from the distribution of CO(2--1) \citep{Oosterloo2024}. Such interactions can influence both the chemical properties of the gas and its dynamical state. In this section we examine several observational diagnostics that help to characterise these effects. We first discuss the HCN/HCO$^{+}$ ratio and evaluate whether radiative transfer effects, in particular high optical depth, may influence its interpretation. We then turn to the enhanced HCO$^{+}$/CO ratio observed at the western edge of the disc and investigate the physical processes that may give rise to this feature.

\subsection{Optical depth effects on the HCN/HCO$^{+}$ ratio}
\label{subsec:opacity_hcn_hcop}

The HCN/HCO$^{+}$ intensity ratio has long been used as a diagnostic of the dominant power source in galaxy nuclei. Studies based on higher-$J$ transitions have shown that enhanced ratios are often associated with AGN-dominated environments \citep[e.g.][]{Izumi2013, Izumi2016, Josh2022, Josh2025}, a result interpreted in terms of both chemistry and excitation. X-ray irradiation and the associated increase in ionisation rate can enhance the HCN abundance relative to HCO$^{+}$, while the dense and warm gas conditions near AGN can also favour the excitation of HCN transitions \citep{Izumi2016, Josh2022}. At the same time, \citet{Josh2025} stressed that intensity ratios are unreliable tracers of the underlying abundances when the lines are optically thick, and that column density ratios provide a more robust diagnostic. As noted in Sect.~\ref{sec:lineimageratio}, the observed HCN(3--2)/HCO$^{+}$(3--2) intensity ratio in 3C\,84 remains close to or below unity across the circumnuclear disc, in apparent tension with the elevated values commonly reported in AGN hosts.
 
Our modelling confirms that this tension is largely an optical depth effect. The optical depth maps derived from the framework (bottom row of Fig.~\ref{fig:physical_parameter_maps}) reveal that HCN(3--2) is frequently optically thick across the circumnuclear disc, with $\tau_{\rm HCN} > 1$ and values reaching $\sim 3$ in several regions (orange highlighted hexagons), while HCO$^{+}$(3--2) and CO(2--1) remain significantly less opaque over most of the disc. When a line becomes optically thick, its intensity saturates and increases only weakly with further increases in column density, so the observed HCN(3--2) emission no longer scales linearly with the underlying HCN abundance. Even if HCN is chemically enhanced relative to HCO$^{+}$, the measured intensity ratio can therefore remain close to unity or below it. Indeed, in the regions where HCN(3--2) is optically thick, the column densities returned by our modelling yield an average abundance ratio of $X({\rm HCN})/X({\rm HCO^{+}}) \gtrsim 3$, confirming that the intensity ratio substantially underestimates the true abundance contrast, in line with the argument of \citet{Josh2025}.
 
Since 3C\,84 is optically classified as a low excitation galaxy \citep{Buttiglione2009, Buttiglione2010}, with a radiative output substantially weaker than in high excitation AGN, X-ray driven chemistry is unlikely to be the dominant mechanism behind this HCN enhancement. To better constrain the physical processes that shape the molecular gas in the disc, including the possible role of the elevated cosmic ray ionisation rates found across the disc (Sect.~\ref{subsection:parametermaps}), it is necessary to consider additional tracers. In the following section we examine the spatial distribution of the HCO$^{+}$/CO ratio, which provides complementary information on the physical state of the molecular gas.

\subsection{Spatial variations of the $HCO^+/CO$ ratio and possible physical origins}
\label{subsection:HCO+/CO discussion}

\begin{figure*}[htbp!]
  \centering
  \includegraphics[width=\textwidth]{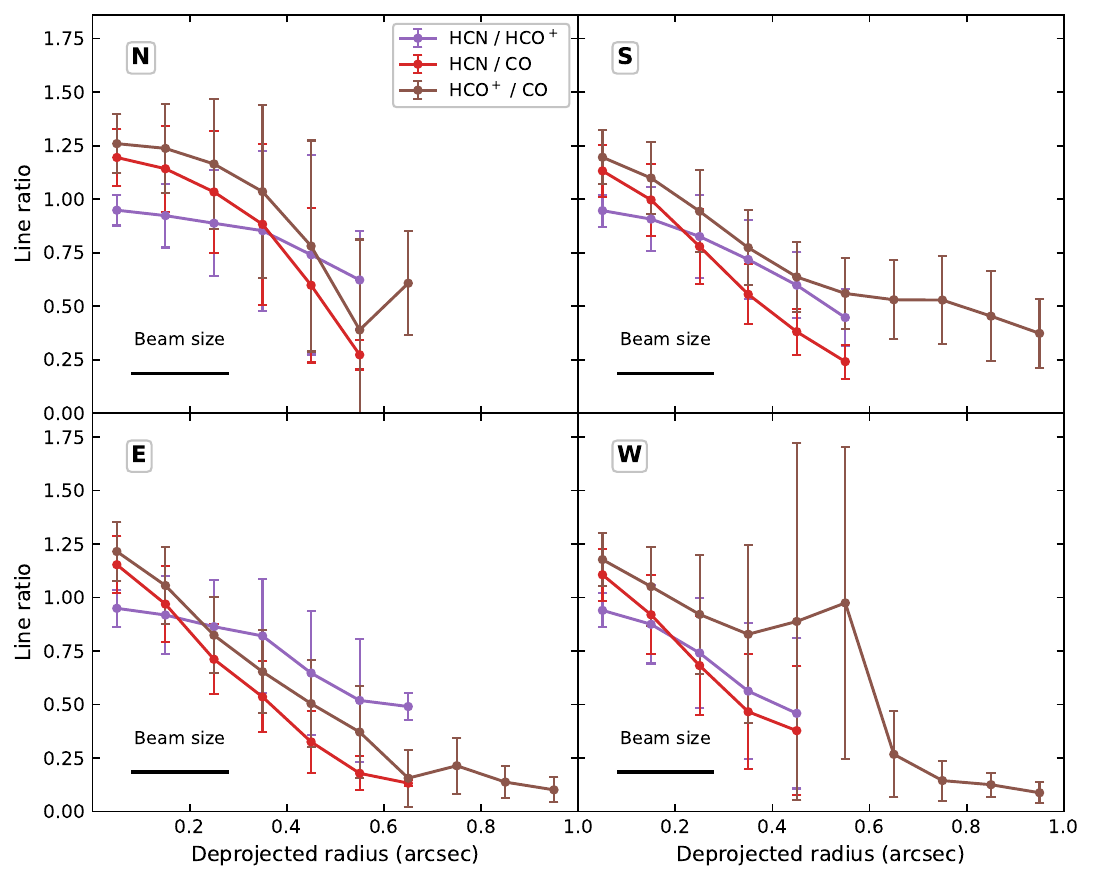}
  \caption{Radial profiles of the $\mathrm{HCN}/\mathrm{HCO}^{+}$,
    $\mathrm{HCN}/\mathrm{CO}$, and $\mathrm{HCO}^{+}/\mathrm{CO}$
    intensity ratios computed in four azimuthal sectors of the deprojected
    disc plane (north, south, east, and west). The disc geometry is defined
    by $\mathrm{PA} = 68^{\circ}$ and $i = 45^{\circ}$. Error bars indicate
    the $1\sigma$ uncertainty within each annulus. The horizontal bar in each
    panel indicates the synthesised beam size.}
  \label{fig:ratio_sector_profiles}
\end{figure*}

The azimuthally averaged radial profiles presented in Sect.~\ref{sec:lineimageratio} show that $\mathrm{HCO}^{+}/\mathrm{CO}$ remains the highest of the three line ratios at all radii. To investigate whether this behaviour is uniform across the disc, we compute radial profiles of the line ratios in four azimuthal sectors using the same deprojected annuli. Figure~\ref{fig:ratio_sector_profiles} presents the resulting profiles. In the northern, eastern, and southern sectors all three ratios decrease with radius, following the global trend. The western sector, however, shows a distinct behaviour: while $\mathrm{HCN}/\mathrm{CO}$ continues to decline, the $\mathrm{HCO}^{+}/\mathrm{CO}$ ratio exhibits a secondary peak around $0\farcs4$--$0\farcs6$, reaching values comparable to those in the inner disc. This feature is not seen in any other sector and indicates a localised enhancement confined to the western boundary of the disc. We note, however, that the uncertainties on this secondary peak are large, reflecting the faintness of the CO emission in this region. 

Because $\mathrm{HCO}^{+}(3{-}2)$ and CO(2--1) arise from different rotational transitions, the absolute values of their ratio also depend on the excitation conditions and should be interpreted with care, although the relative variation across sectors is robust since any systematic scaling between the two transitions affects all directions equally.

To assess whether the western enhancement persists when accounting for the excitation mismatch between the two transitions, we construct a proxy CO(3--2) map by scaling the observed CO(2--1) intensity. The CO spectral line energy distribution of 3C~84 shows that the emission is increasingly populated toward higher-$J$ transitions \citep{Esposito2024}, implying that CO(2--1) underestimates the CO(3--2) intensity. We adopt two bracketing scaling factors: $R_{32/21} = 1.5$ from the observed global brightness temperature ratio \citep{Esposito2024}, and $R_{32/21} = 2.07$ from the upper bound of CO excitation models computed with \textsc{SpectralRadex} for conditions representative of the circumnuclear environment (see Appendix~\ref{sec:proxy3-2} for the resulting proxy maps). Even under the most conservative correction, which maximises the estimated CO(3--2) intensity, the $\mathrm{HCO}^{+}(3{-}2)/\mathrm{CO}(3{-}2)$ ratio remains above unity along the western edge of the disc. The localised nature of this enhancement, confined to the western boundary and absent in the other sectors, raises the question of what physical process is responsible.

From a morphological perspective, the enhanced $\mathrm{HCO}^{+}/\mathrm{CO}$ structure appears as a narrow layer along the western boundary of the disc, consistent with the structure expected in photodissociation regions (PDRs; \citealt{Wolfire2022}). In such UV-irradiated environments, $\mathrm{HCO}^{+}$ production is enhanced through warm carbon chemistry at the dissociation front, while CO remains suppressed by photodissociation until sufficient column density allows self-shielding \citep{Visser2009, Goicoechea2016, Wolfire2022}. If such a PDR scenario applies to the western edge of the disc, a sufficient ultraviolet radiation field must be present at distances of $200$--$300$~pc from the nucleus. \citet{Punsly2018} reported a monochromatic luminosity of $\lambda L_{\lambda}(1450\,\text{\AA}) \approx 3.5 \times 10^{42}\,\mathrm{erg\,s^{-1}}$ for the accretion flow\footnote{Adopting a power-law continuum $F_{\lambda} \propto \lambda^{-1.5}$ from \citet{Punsly2018}, integration over the standard FUV band (912--2400\,\AA) gives $L_{\mathrm{FUV}} \approx 3 \times 10^{42}\,\mathrm{erg\,s^{-1}}$, corresponding to $F_{\mathrm{FUV}} \approx 0.27\,\mathrm{erg\,cm^{-2}\,s^{-1}}$ or $G_{0} \approx 1.7 \times 10^{2}$ at 300~pc.}, indicating that the AGN provides a moderate FUV field at these radii. Additional UV photons from local star formation at a rate of $\sim 3\,M_{\odot}\,\mathrm{yr}^{-1}$ \citep{Silver1998} can further sustain PDR chemistry in the outer disc.

A PDR origin alone, however, does not fully explain why the enhancement is confined to the western side of the disc. The AGN radiation field is expected to be roughly isotropic at these scales. The location of the enhanced ratio instead coincides with the region where large-scale molecular filaments connect to the circumnuclear disc. High-resolution CO(2--1) observations show that several filamentary structures converge toward the nucleus and supply cold gas to the rotating disc \citep{Oosterloo2024}, and the western side corresponds to one of the main accretion interfaces. The filamentary gas approaching from the west is predominantly redshifted at $v \sim +50$ to $+100$\,km\,s$^{-1}$ relative to systemic, while the adjacent western side of the disc is blueshifted at $v \sim -100$ to $-150$\,km\,s$^{-1}$ \citep{Oosterloo2024}, producing a line-of-sight velocity difference of $\sim 150$--$250$\,km\,s$^{-1}$ at the interface. Since these are projected velocities, the true three-dimensional shear is likely larger. The resulting velocity shear and compression can drive large-scale shocks at the boundary between the infalling gas and the rotating disc \citep{Banda2016, Sormani2019}. At the shock front the gas is compressed, increasing the fraction of dense material and thereby boosting the excitation of high critical density tracers such as $\mathrm{HCO}^{+}$. The sputtering of ice mantles and the generation of a local UV radiation field at the shock interface can further trigger a transient chemistry in which $\mathrm{HCO}^{+}$ is efficiently produced through reactions of C$^{+}$ with the freshly released molecular material \citep{Rawlings2000}. In addition, the electron density enhancement predicted within the magnetic precursor of C-shocks increases the excitation of ionic species through electron impact collisions \citep{Roberts2010}, an effect that is particularly relevant for $\mathrm{HCO}^{+}$ given its sensitivity to the local electron abundance. Meanwhile, CO emission can be suppressed in shocked environments through dissociation or excitation effects \citep{Godard2019, Zhou2022, Tu2024}. The combination of enhanced $\mathrm{HCO}^{+}$ production and excitation with reduced CO emission provides a natural explanation for the localised $\mathrm{HCO}^{+}/\mathrm{CO}$ enhancement observed in the western sector, and the azimuthal asymmetry that a PDR scenario alone cannot account for. We regard this shock interpretation as plausible rather than established, since the CO emission is faint along the western edge and the current data do not resolve the kinematic signature of a shock at the interface.

This study focuses on interpreting the line intensities, and a full kinematic analysis is beyond its scope, especially as the kinematics of the CND have already been studied in detail by \citet{Oosterloo2024}. The line profiles still carry information about the velocity shear at the western interface, but at the 72\,pc resolution of these data the fast inner rotation and the interface are unresolved, so beam smearing dominates the line widths and velocity dispersion. We therefore summarise the kinematic evidence that the current data can support in Appendix~\ref{app:kinematics}, where the western hexagons sit at the disc velocity with somewhat larger line widths than elsewhere at the same radius.

\section{Conclusions}
\label{section:conclusion}
We have presented a spatially resolved analysis of the molecular gas in the circumnuclear disc of 3C~84, combining ALMA observations of CO(2--1), HCN(3--2), and HCO$^+$(3--2) at a resolution of $\sim 72$\,pc with a Bayesian inference framework that couples time-dependent chemistry and non-LTE radiative transfer through a neural network emulator. The main results are summarised as follows.
 
\begin{enumerate}
 
\item The integrated intensity maps show that all three molecular transitions trace the circumnuclear disc, with CO(2--1) being the most extended component and the dense gas tracers HCN(3--2) and HCO$^+$(3--2) more centrally concentrated. HCO$^+$(3--2) extends farther than HCN(3--2) in the outer disc, consistent with its lower critical density. All three line ratios peak in the inner disc, decline with radius, and reach values significantly higher than in nearby star-forming galaxies, comparable to those found in AGN-dominated nuclei.
 
\item The Bayesian analysis results indicate radial gradients. The gas density peaks at $\log_{10}\,n(\mathrm{H_2}) \approx 6.3$ in the inner disc and decreases to $\sim 5$ at larger radii, while the kinetic temperature drops from $T_\mathrm{gas} \sim 200$\,K to $\sim 160$\,K over the same range. The cosmic ray ionisation rate is strongly enhanced throughout the disc, from $\log_{10}(\zeta/\zeta_0) \approx 4.9$ in the centre to $\sim 3$ at larger radii, with a notable exception on the western side where locally elevated values of $\log_{10}(\zeta/\zeta_0) \gtrsim 4$ may be related to the interaction between infalling filaments and the rotating disc. These radial trends persist when the adopted column density is varied, while the absolute parameter values for individual regions remain model dependent.
 
\item Despite the presence of a powerful radio AGN, the observed HCN(3--2)/HCO$^+$(3--2) intensity ratio remains close to or below unity across the disc. Our modelling shows that HCN(3--2) is frequently optically thick ($\tau \sim 1$--3), while HCO$^+$(3--2) and CO(2--1) remain less opaque. The saturation of the HCN line suppresses the observed ratio even though the underlying abundance ratio $X({\rm HCN})/X({\rm HCO^{+}}) \gtrsim 3$ in the optically thick regions, demonstrating that the HCN/HCO$^+$ intensity ratio cannot be used as a direct abundance diagnostic without accounting for optical depth \citep{Josh2025}.
 
\item Azimuthally resolved radial profiles show a localised enhancement of HCO$^+$/CO at the western boundary of the disc, absent in the other sectors and persistent after correcting for the excitation mismatch between HCO$^+$(3--2) and CO(2--1). This feature coincides spatially with the region where large-scale molecular filaments connect to the rotating disc \citep{Oosterloo2024} and is consistent with shock processing at the filament--disc interface enhancing HCO$^+$ production while suppressing CO emission. Because the CO emission is faint here, we treat this shock scenario as a plausible origin rather than a confirmed one. A photodissociation region origin may contribute but cannot alone account for the azimuthal asymmetry. 

 
\end{enumerate}

These results suggest that the circumnuclear disc of 3C~84 is not a dynamically or chemically isolated structure, but is actively shaped by the accretion of cold gas from the surrounding filamentary environment. Future observations are essential to further isolate the dominant excitation mechanism in the western region. High resolution measurements of the CO(3--2) transition would allow the HCO$^+$(3--2)/CO(3--2) intensity ratio to be determined directly, eliminating the uncertainties associated with the proxy conversion. Furthermore, targeted observations of classical shock tracers, such as SiO, HNCO, or CH$_3$OH, would provide critical constraints on the extent of mechanical processing. The combination of spatially resolved line ratios, Bayesian physical parameter estimation, and optical depth modelling presented in this work provides a framework that can be applied to other AGN circumnuclear environments where the interplay between feeding, feedback, and local processing determines the conditions of the molecular gas.

\begin{acknowledgements}
We thank the anonymous referee for their comments and constructive suggestions which improved this manuscript. BJ acknowledges support from the CSC (China Scholarship Council) scholarship program. SV acknowledges support by the European Research Council (ERC) Advanced Grant MOPPEX 833460.vii.
\end{acknowledgements}

\bibliographystyle{aa}
\bibliography{main}

\begin{appendix}

\section{Improvement of data quality}
\label{sec:appData}

\begin{figure}[h!]  
\centering
    \includegraphics[width=0.33\textwidth]{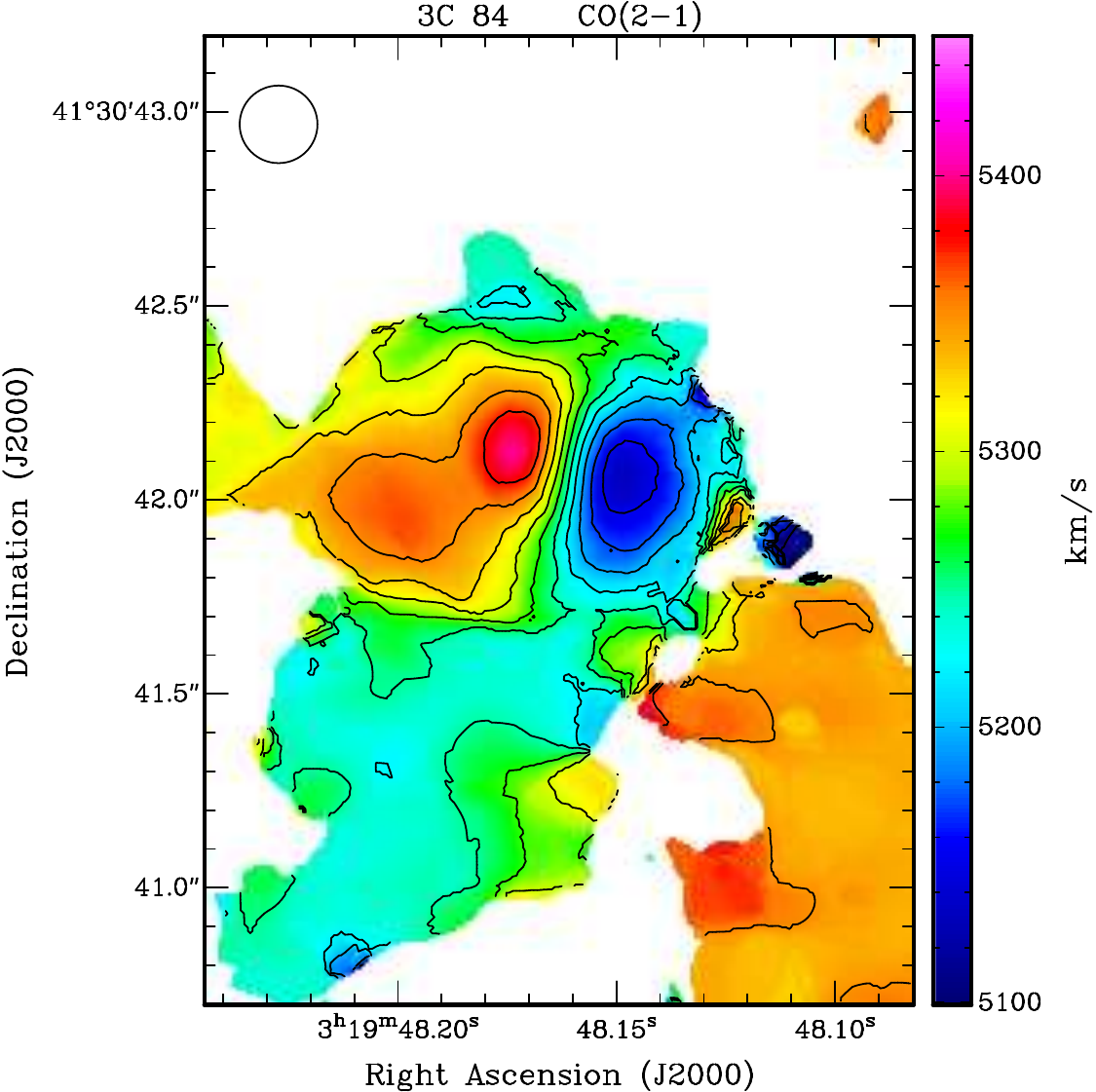}
    \vskip0.3cm
    \includegraphics[width=0.33\textwidth]{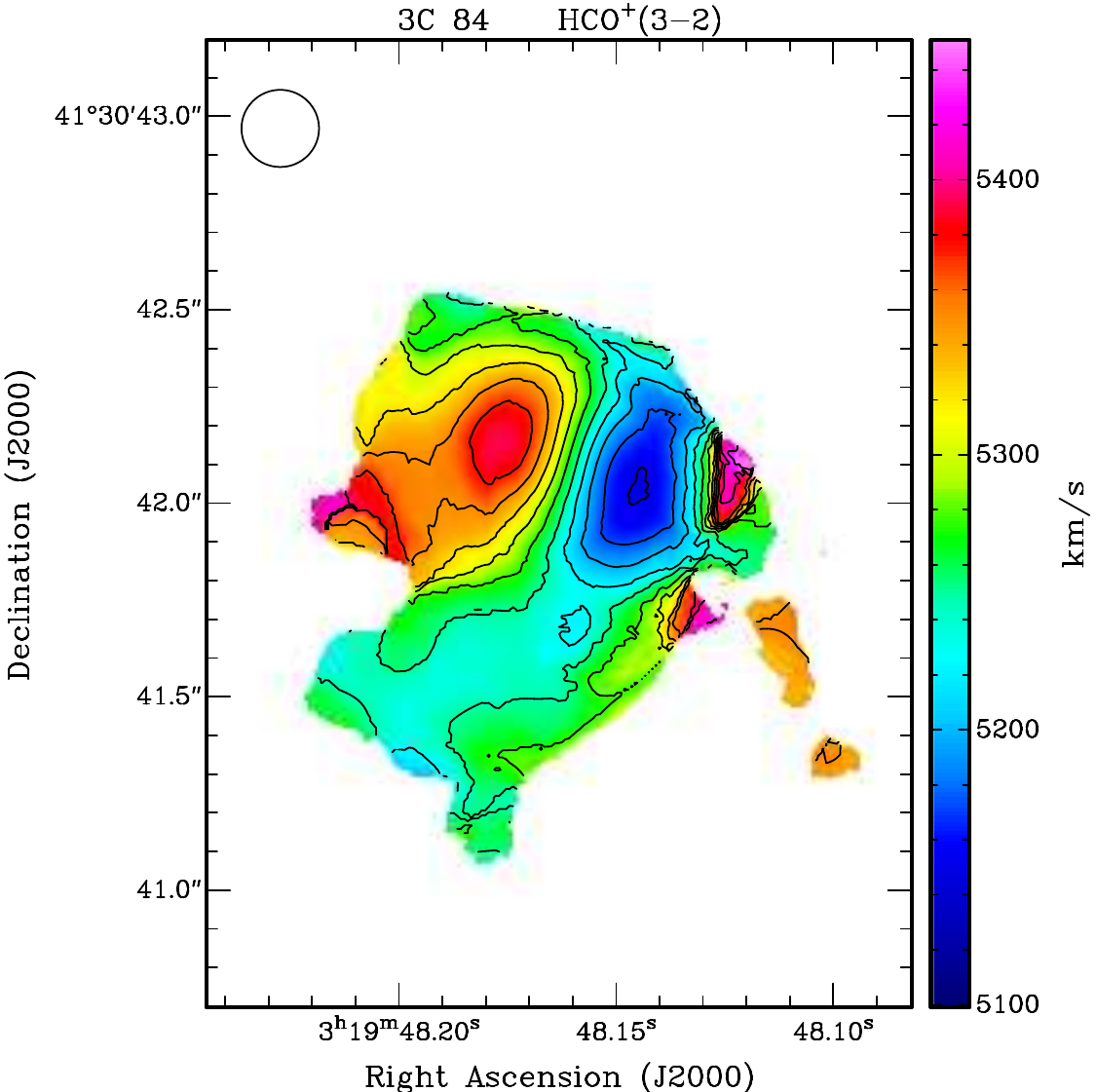}
    \vskip0.3cm
    \includegraphics[width=0.33\textwidth]{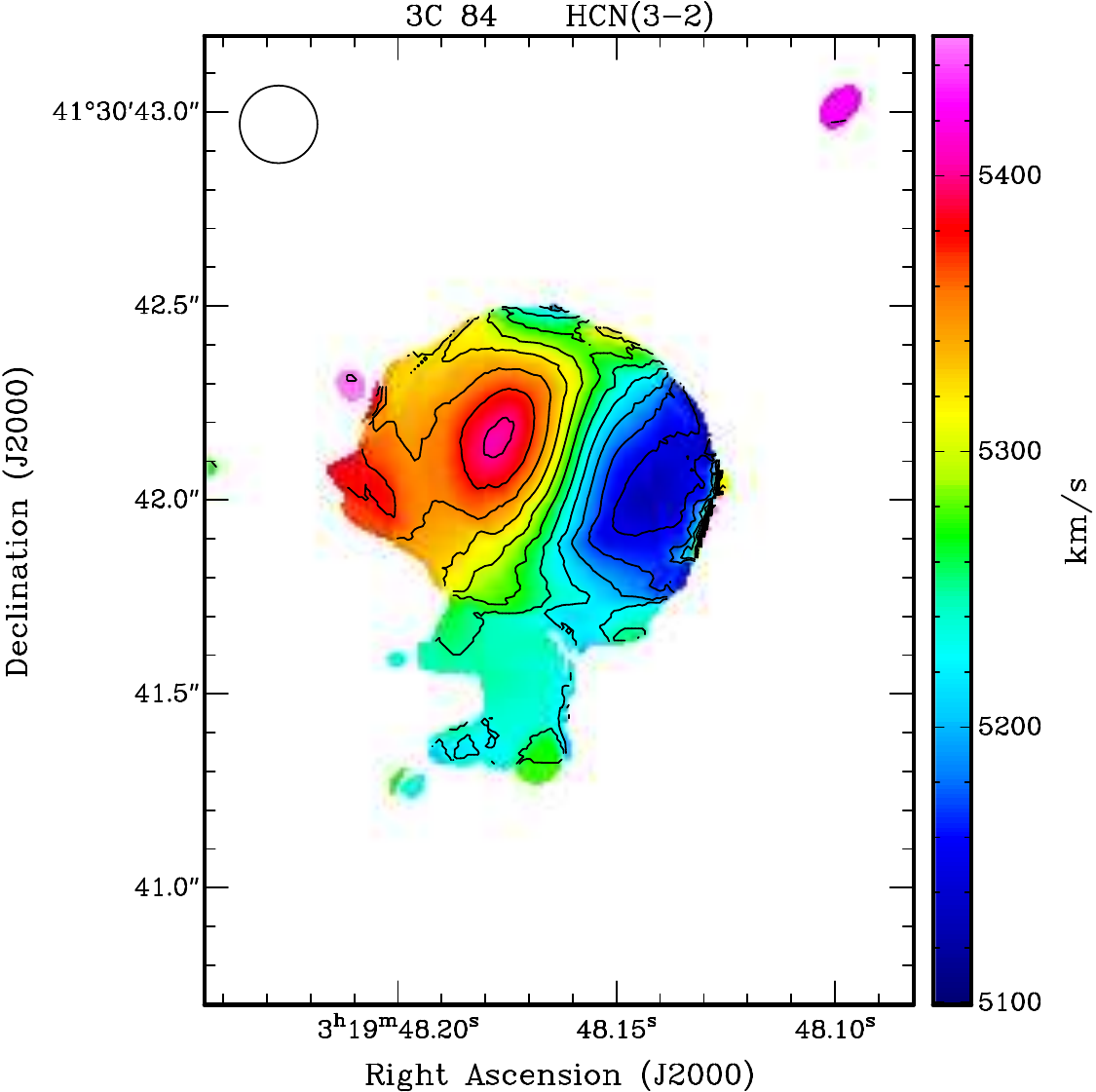}
    \caption{Velocity fields of the CO(2--1) (top), HCO$^+$(3--2) (middle), and HCN(3--2) (bottom) emission in and around 3C 84. Contour levels are 5100, 5125, 5150, ...  km s$^{-1}$. The beam is indicated in the top-left corner of each panel.}
    \label{fig:velfies}
\end{figure}

\begin{figure}[b!]
    \centering
    \includegraphics[width=0.33\textwidth]{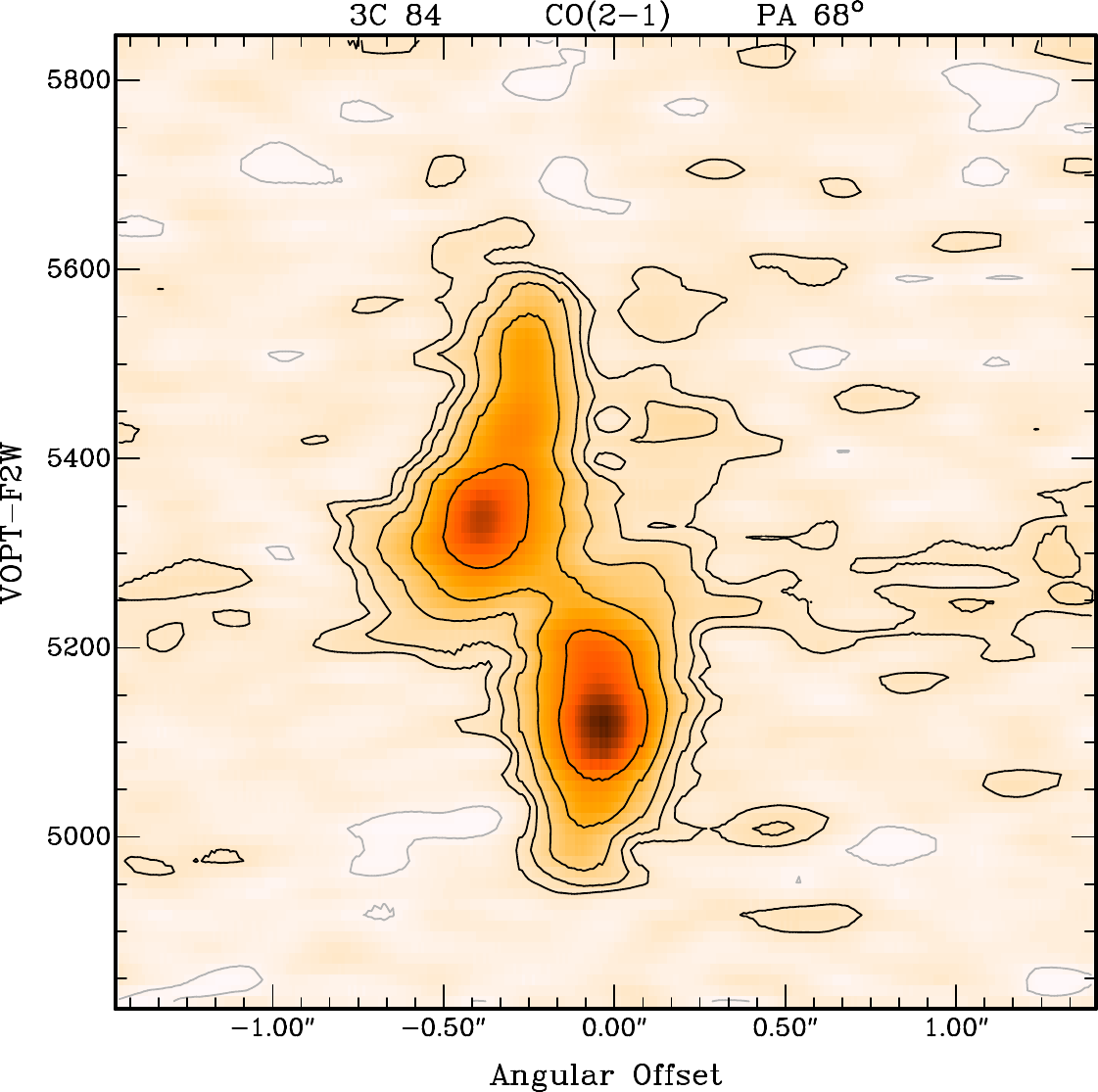}
    \vskip0.3cm
    \includegraphics[width=0.33\textwidth]{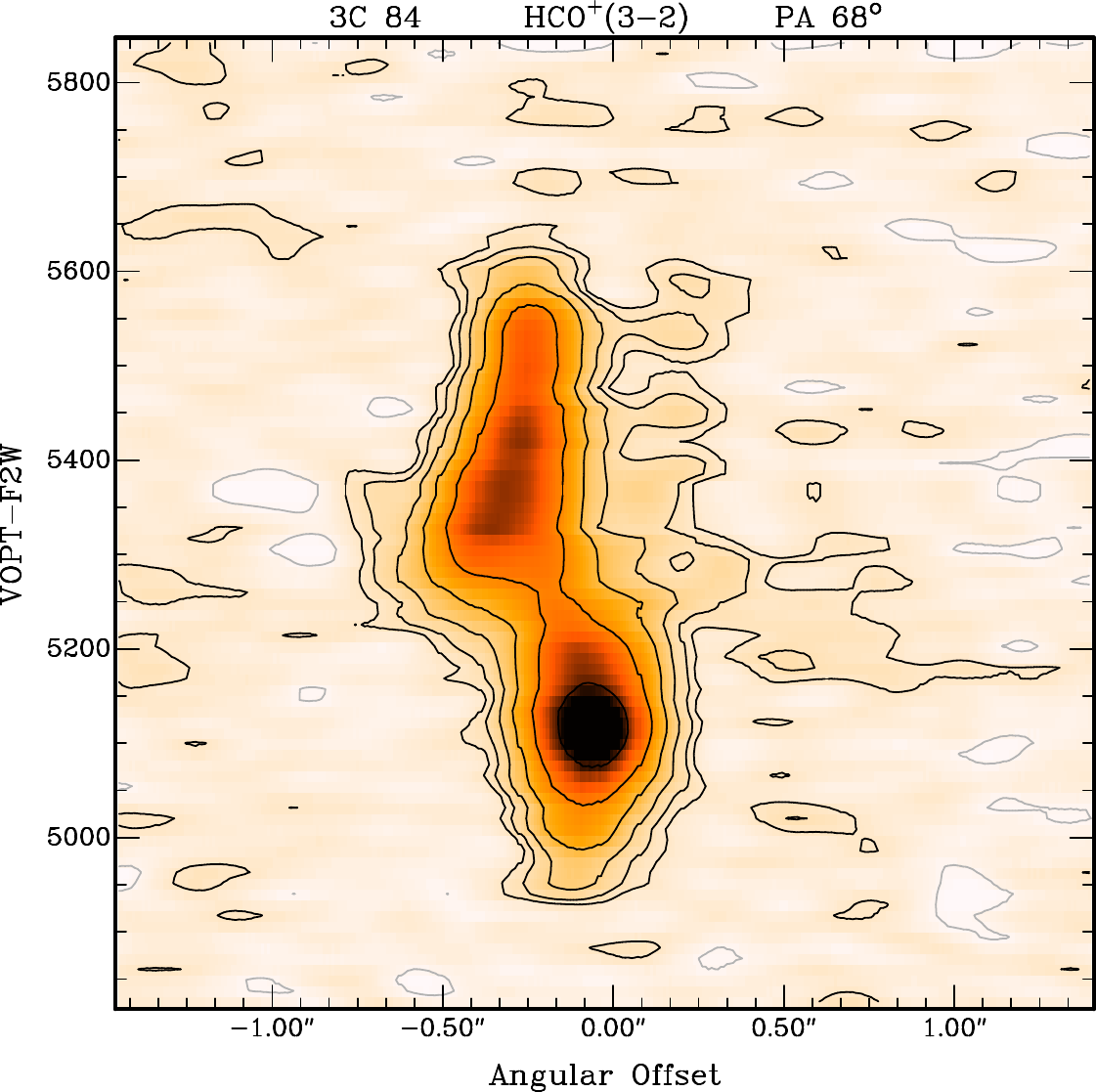}
    \vskip0.3cm
    \includegraphics[width=0.33\textwidth]{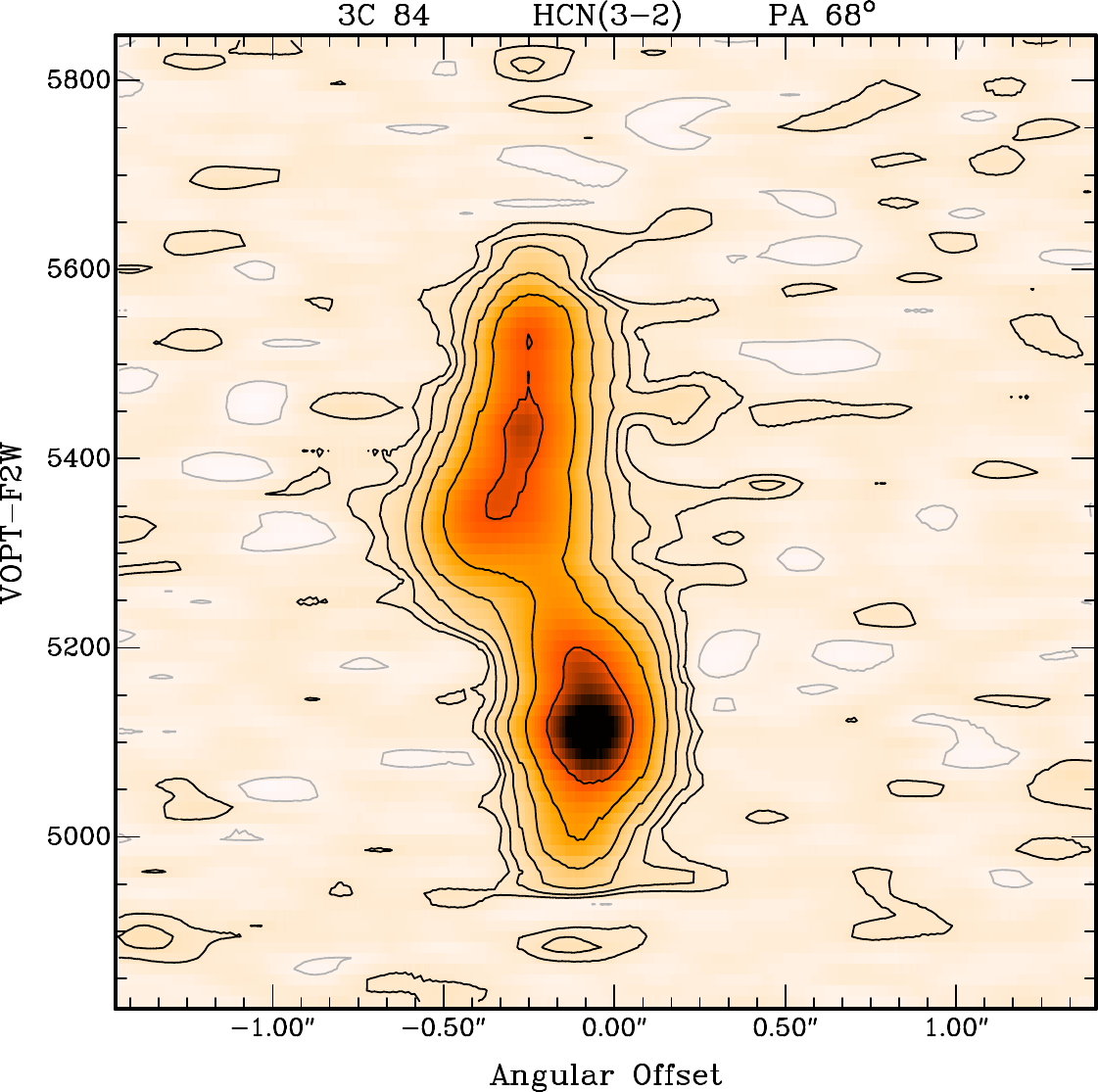}
    \caption{Position-velocity plots taken along position angle 68$^\circ$ of the CO(2--1) (top), HCO$^+$(3--2) (middle), and HCN(3--2) (bottom) emission of the circumnuclear disk. The  lowest contour level is 1.5$\sigma$ and subsequent levels increase with a factor 2.}
    \label{fig:slices}
\end{figure}

To illustrate the improved data quality due to the spectral self calibration we applied to the archival data, in Fig.\ \ref{fig:velfies} we show the velocity fields of the three emission lines observed, and in Fig.\ \ref{fig:slices} for the  three species the position-velocity plots  of the CND taken along position angle 68$^\circ$. These plots should be compared with Figs 1, 2, 5, and 6 of \citet{Nagai2019} which show data products of the same observation but produced by the standard ALMA pipeline.

\clearpage

\section{$N_{H_{2}}$ sensitivity check}
\label{sec:appx_Nh2check}

Although we reduced the dimensionality of the Bayesian inference, we acknowledge that in the first stage, where we performed the $\chi^{2}$-fit, three observed line intensities were used to constrain four parameters. To assess the sensitivity of our results to the adopted $N_{\mathrm{tot}}$, we explored a range of $\log_{10} N_{\mathrm{H_2}}$ values around the best-fit value. We selected the two regions shown in Fig.~\ref{fig:corner_examples} as representative examples for this test.

\begin{figure*}[h!]
\centering
\subfigure{\includegraphics[width=0.3\textwidth]{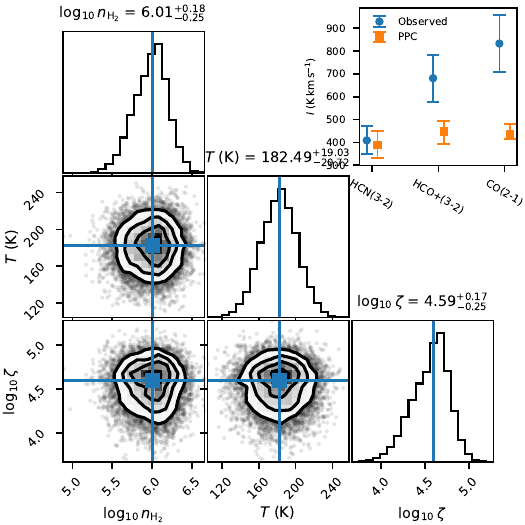}\label{fig:a}}
\subfigure{\includegraphics[width=0.3\textwidth]{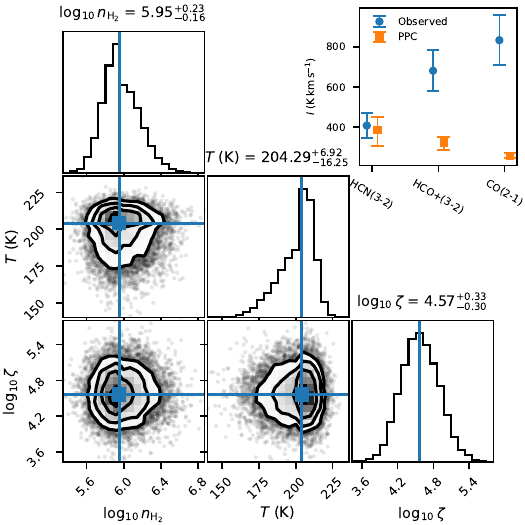}\label{fig:b}}
\subfigure{\includegraphics[width=0.3\textwidth]{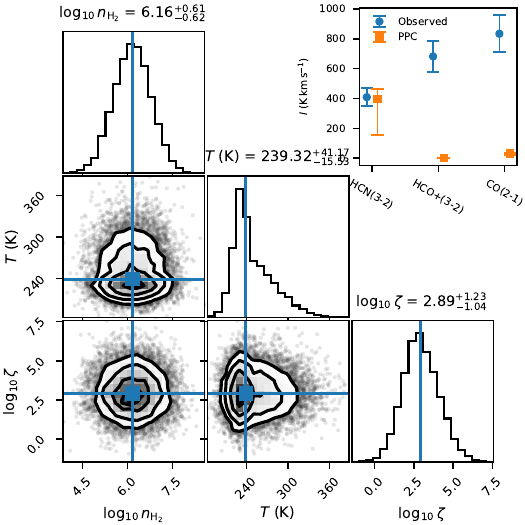}\label{fig:c}}
\subfigure{\includegraphics[width=0.3\textwidth]{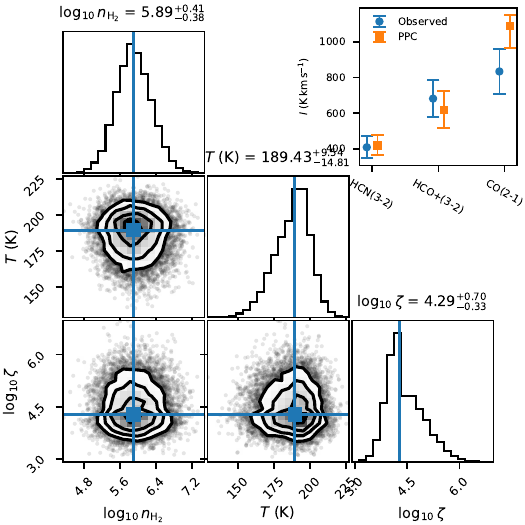}\label{fig:d}}
\subfigure{\includegraphics[width=0.3\textwidth]{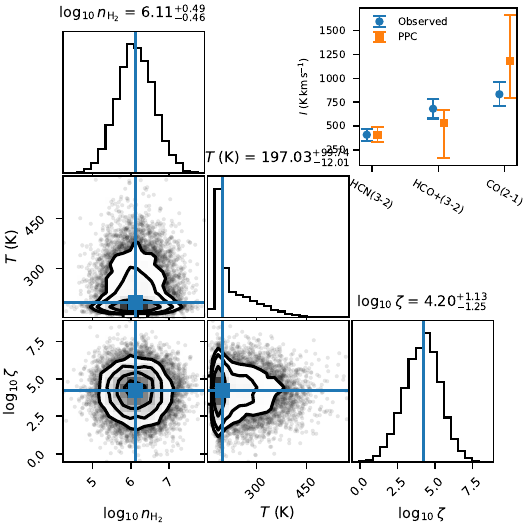}\label{fig:e}}
\subfigure{\includegraphics[width=0.3\textwidth]{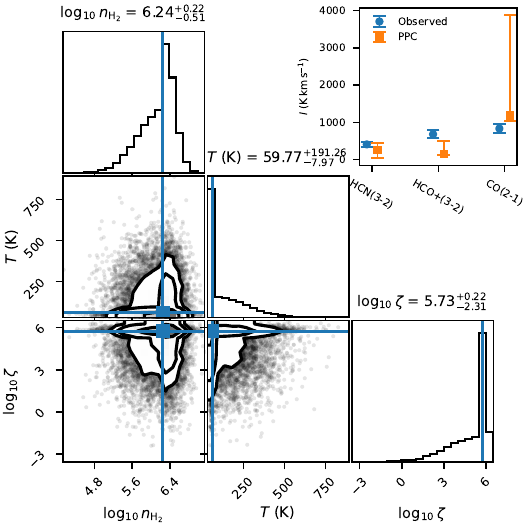}\label{fig:f}}
\caption{Sensitivity to variations in $N_{\mathrm{tot}}$ for the region highlighted in yellow in Fig.~\ref{fig:physical_parameter_maps}, for which the best-fit column density is $\log_{10}(N_{\mathrm{H_2}}/\mathrm{cm^{-2}}) = 22.60 \pm 0.19$. The top row shows the results when $N_{\mathrm{tot}}$ is decreased by 0.2, 0.4, and 1.2~dex from the best-fitted value ($\log_{10} N_{\mathrm{H_2}} = 22.40$, 22.20, and 21.40), while the bottom row shows the corresponding increases ($\log_{10} N_{\mathrm{H_2}} = 22.80$, 23.00, and 23.80). In the top row the predicted HCO$^+$(3--2) and CO(2--1) intensities fall below the observed values by more than $1\sigma$, so none of these cases meets the acceptance criterion of the $N_{\mathrm{tot}}$ test (all three predicted intensities within $1\sigma$ of the observed values). In the bottom row only the $+0.4$~dex case meets it and is retained in Fig.~\ref{fig:galleryofNtotcheck}. The $\pm1.2$~dex cases lie outside the range of the test and are shown to illustrate the behaviour at large offsets.}
\label{fig:grid}
\end{figure*}

\begin{figure*}
\centering
\subfigure{\includegraphics[width=0.3\textwidth]{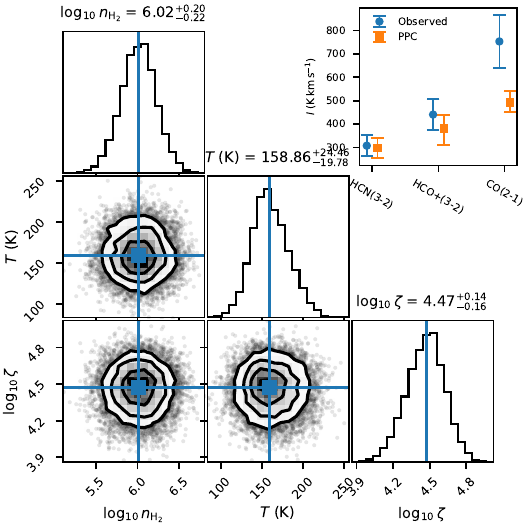}\label{fig:a}}
\subfigure{\includegraphics[width=0.3\textwidth]{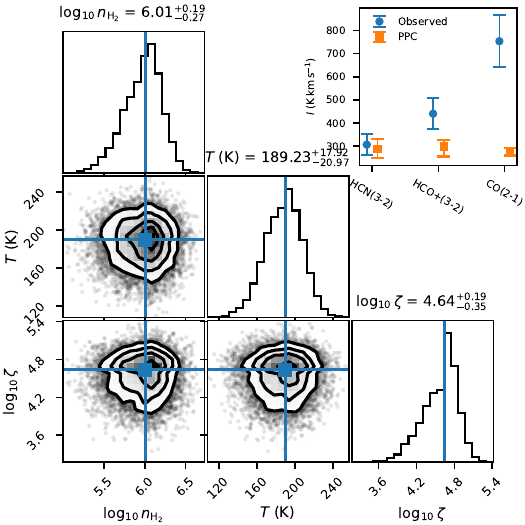}\label{fig:b}}
\subfigure{\includegraphics[width=0.3\textwidth]{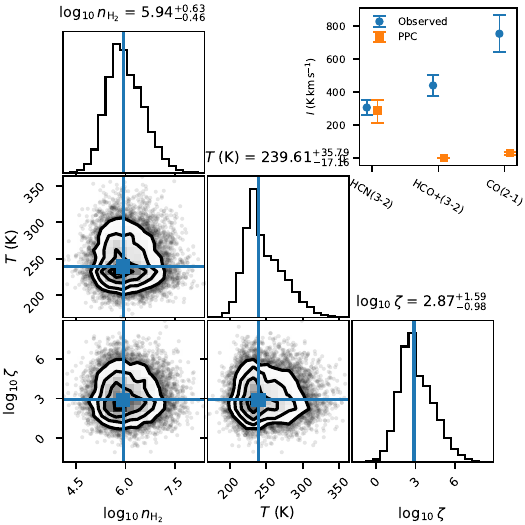}\label{fig:c}}
\subfigure{\includegraphics[width=0.3\textwidth]{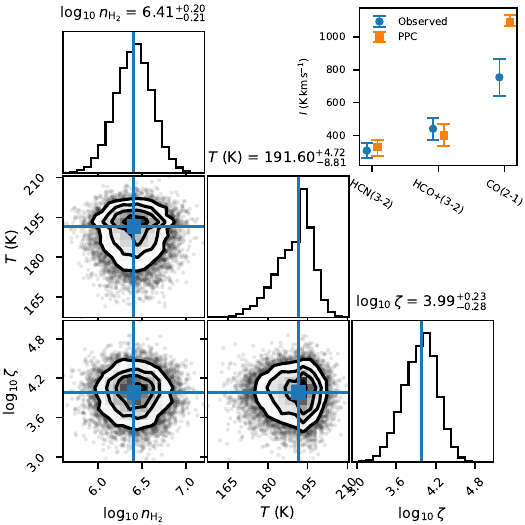}\label{fig:d}}
\subfigure{\includegraphics[width=0.3\textwidth]{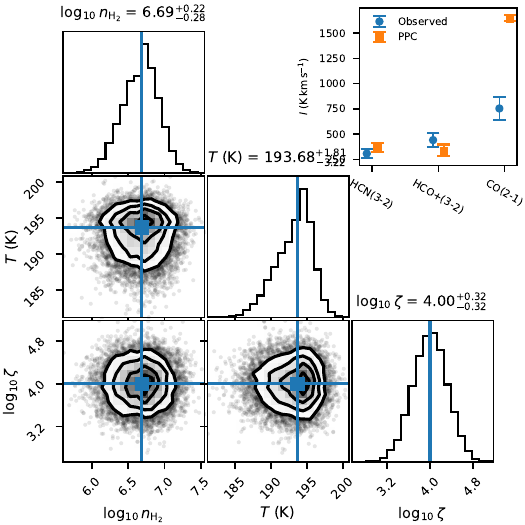}\label{fig:e}}
\subfigure{\includegraphics[width=0.3\textwidth]{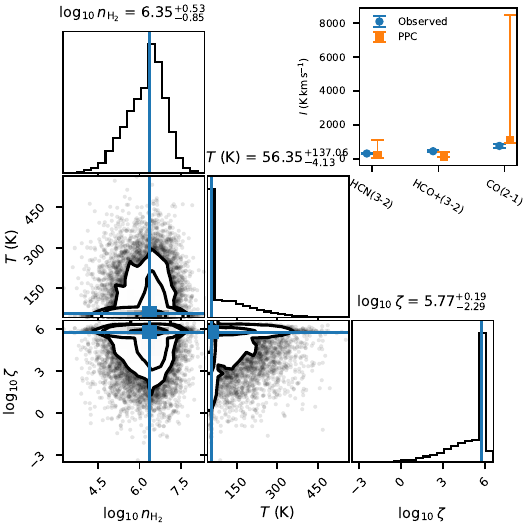}\label{fig:f}}
\caption{Same as Fig.~\ref{fig:grid}, but for the region highlighted in magenta in Fig.~\ref{fig:physical_parameter_maps}, for which the best-fit column density is $\log_{10}(N_{\rm H_2}/\mathrm{cm^{-2}}) = 22.67 \pm 0.15$. None of the variations shown is retained under the acceptance criterion, so this region is absent from Fig.~\ref{fig:galleryofNtotcheck}}.
\label{fig:grid_B}
\end{figure*}

In Fig.~\ref{fig:grid} and Fig.~\ref{fig:grid_B}, we present the posterior corner plots together with the PPCs for these two regions. Fig.~\ref{fig:grid} corresponds to the left panel of Fig.~\ref{fig:corner_examples}, where the top row shows the results when $N_{\mathrm{tot}}$ is decreased by 0.2, 0.4, and 1.2~dex from the best-fit value, while the bottom row shows the corresponding increases. Fig.~\ref{fig:grid_B} corresponds to the right panel of Fig.~\ref{fig:corner_examples}.

For both regions, even a deviation of 0.2~dex in $N_{\mathrm{tot}}$ leads to noticeable discrepancies between the PPCs and the observed intensities, despite the corner plots still appearing reasonable. As the offset increases further, the PPCs deviate more significantly from the observations, and the corner plots reveal that $T_{\mathrm{gas}}$ or $\zeta$ begin to approach the boundaries of the parameter space.

The two examples above show that the inferred parameters of an individual region depend on the adopted column density. To test whether the radial trends reported in Sect.~\ref{subsection:parametermaps} are driven by this choice rather than by the data, we extended the check to all regions with clean detections in the three lines and propagated the $N_{\rm tot}$ variation into the radial profiles of the three parameters. For each region we repeated the second stage of the Bayesian inference after shifting $\log_{10} N_{\rm H_2}$ by $\pm0.2$ and $\pm0.4$ dex from its best-fit value, with all other settings unchanged. We then applied a single acceptance criterion to each shifted case, based on the PPC. A variation is retained only if the predicted intensities of all three lines still overlap the observed values within $1\sigma$. This criterion is motivated by the behaviour of the two example regions described above, in which a shift of $0.2$ dex already moves the predicted intensities away from the observations. Since even moderate shifts often break this agreement, the data themselves limit how far $N_{\rm tot}$ can move, so that the column density is in practice constrained at roughly the $0.2$ dex level. The accepted variations therefore define the systematic range of $N_{\rm tot}$ allowed by the data in each region.

Fig.~\ref{fig:ntot_radial} shows the resulting profiles of gas density, kinetic temperature, and cosmic ray ionisation rate. The faint grey points mark the fiducial value of each hexagon at its deprojected radius. For single-component hexagons, the attached grey bar indicates the full range spanned by that hexagon across its accepted $N_{\rm tot}$ variations. Hexagons that the GMM fit resolves into two density components are instead shown as their two cluster means, plotted as open diamonds joined by a thin line, with each mean treated as a separate fiducial value. The coloured line is the binned radial profile. Its marker gives the mean of the fiducial values in each annulus, and the coloured bar on each marker spans the combined range of the fiducial values of all hexagons in that annulus together with their accepted $N_{\rm tot}$ variations.

In all three panels the fiducial values decline toward larger radii, although the scatter around this trend has a different origin for each parameter. For the gas density, the large scatter is mostly produced by the GMM fit. As illustrated in Fig.~\ref{fig:corner_examples}, the density posterior can separate into a low and a high component. This may reflect the presence of two distinct gas components within the beam, or a genuine degeneracy in the inference. Given that only three line intensities are available, we regard this pattern as a limitation of the current dataset in breaking such degeneracies rather than as a physical result. For the kinetic temperature and the cosmic ray ionisation rate, the scatter in Fig.~\ref{fig:ntot_radial} is instead dominated by the $N_{\rm tot}$ variations, indicated by the range bars. To assess these cases we show in Fig.~\ref{fig:galleryofNtotcheck} all corner plots for regions which pass the PPC after shifting the $N_{\rm tot}$. Some of these accepted cases return high temperatures or ionisation rates, but their posteriors are poorly behaved. For example, region 1 at $+0.2$ dex gives $T = 215.3^{+0.7}_{-1.7}$ K, where the temperature posterior collapses into a narrow spike rather than a resolved peak, and region 23 at $+0.4$ dex gives $\log_{10}(\zeta/\zeta_0) = 5.31^{+0.40}_{-2.49}$, with the distribution piling up against the upper edge of the prior range. In both cases the PPC passes only because the forward model can still reproduce the three observed intensities at these boundary values, so these high values reflect the assumed $N_{\rm tot}$ offset rather than a genuine constraint and should be interpreted with additional caution. Despite the overall decline seen in all three panels of Fig.~\ref{fig:ntot_radial}, the cosmic ray ionisation rate shows a small secondary rise at the largest radii, consistent with the locally elevated values discussed in Sect.~\ref{subsection:parametermaps}.

\begin{figure*}[p]
\centering
\includegraphics[width=\textwidth]{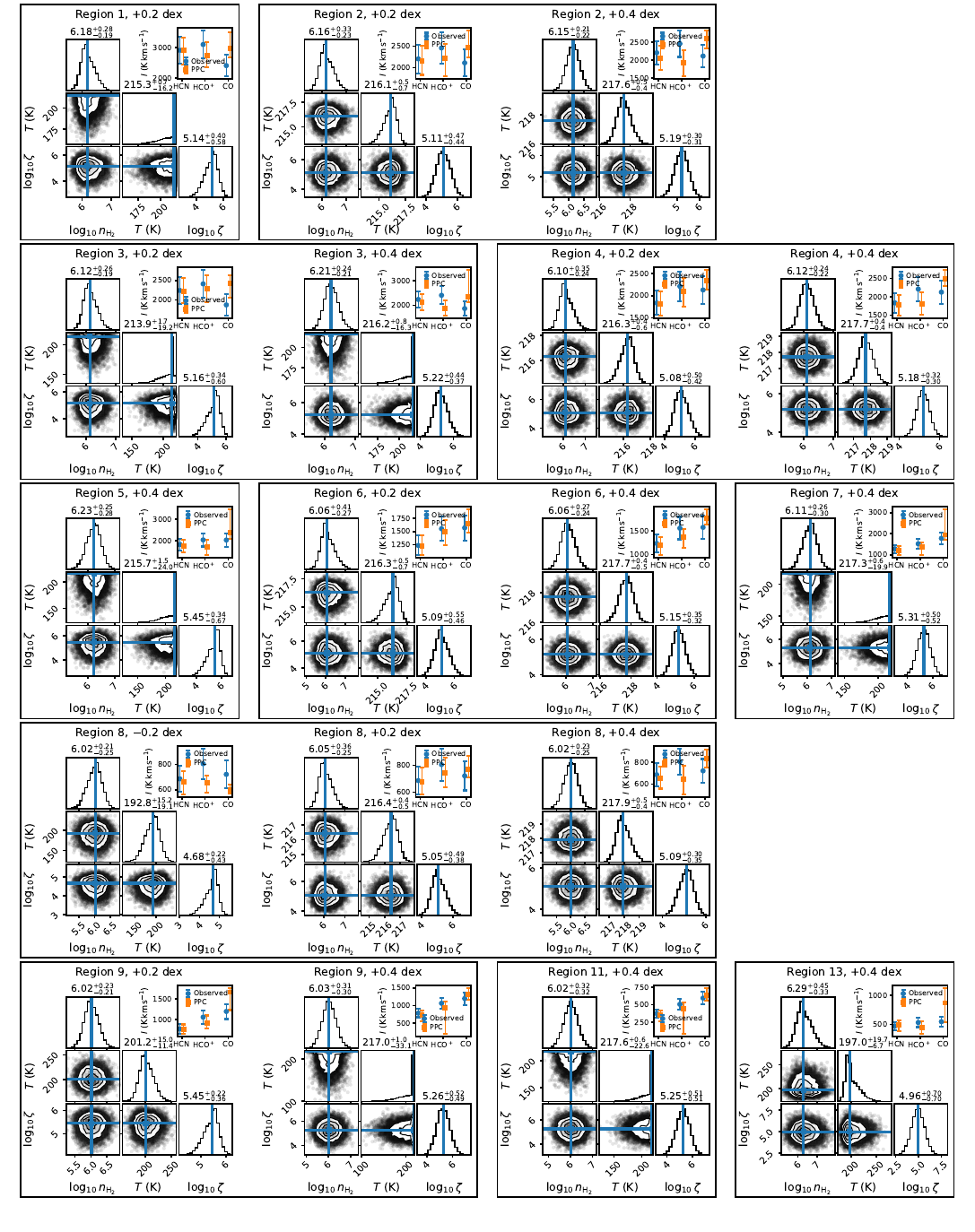}
\end{figure*}

\begin{figure*}[p]
\centering
\includegraphics[width=\textwidth]{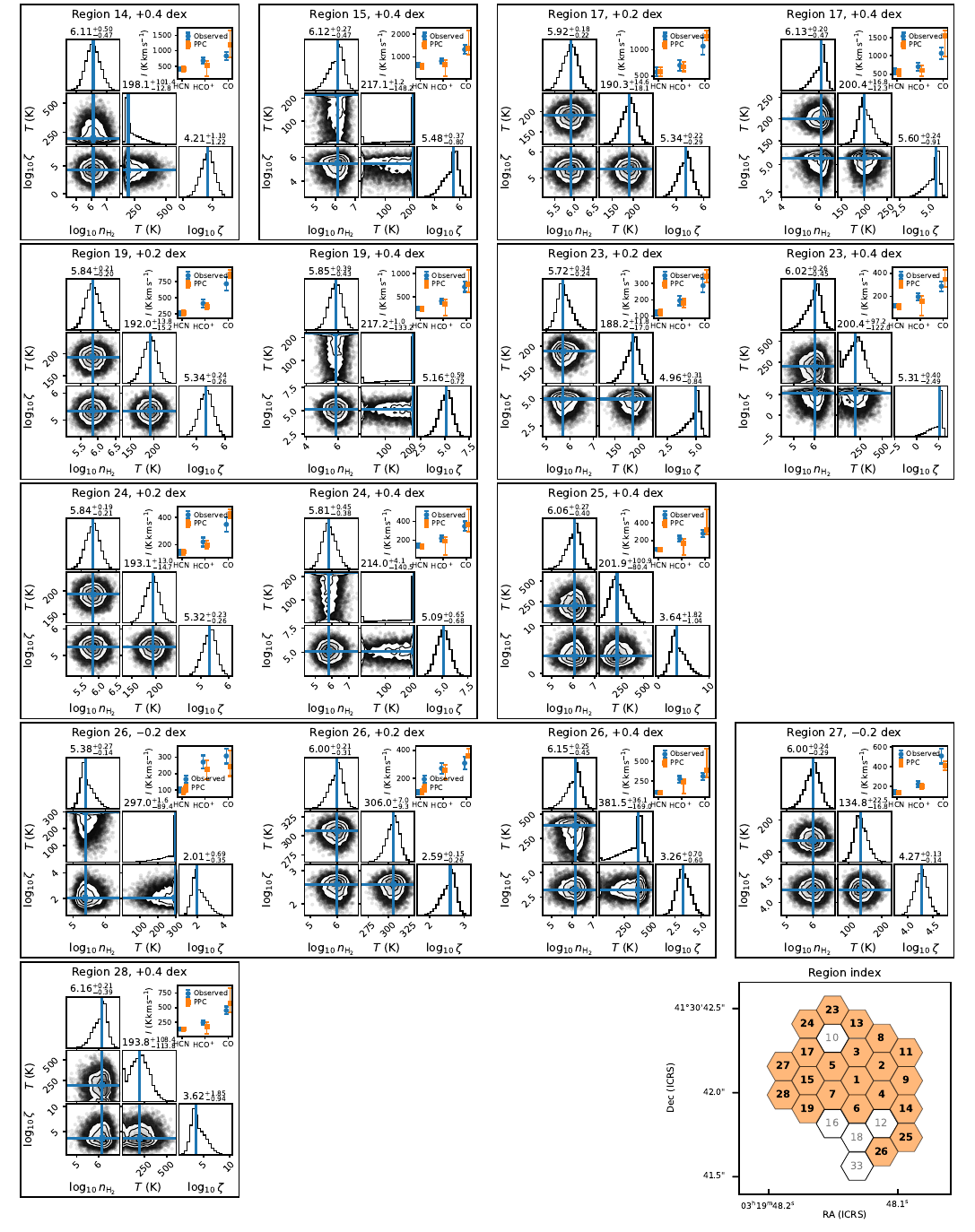}
\caption{Corner plots for regions passing the PPC test after varying $N_{\rm tot}$.}
\label{fig:galleryofNtotcheck}
\end{figure*}

\begin{figure*}
\centering
\includegraphics[width=\textwidth]{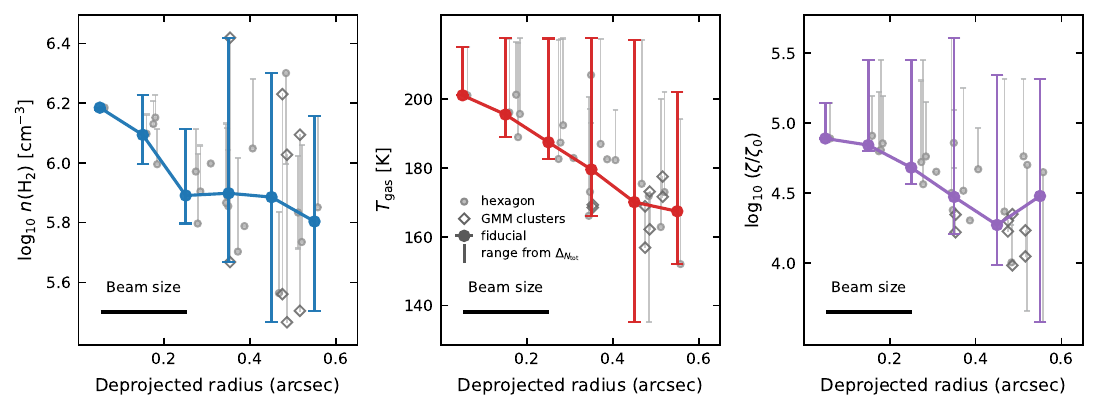}
\caption{Radial profiles of the inferred gas density (left), kinetic temperature (middle), and cosmic ray ionisation rate (right), including the systematic effect of the adopted column density. Faint grey points show the fiducial value of each hexagon at its deprojected radius, with grey bars giving the range covered when $N_{\rm tot}$ is varied by $\pm0.2$ and $\pm0.4$ dex over the cases that remain consistent with the posterior predictive check. The coloured line is the binned radial profile; the marker is the mean of the fiducial values in each annulus and the coloured bar is the combined range of the fiducial values and the accepted variations of all hexagons in that annulus. Only regions detected in all three lines are shown. The black bar indicates the beam size.}
\label{fig:ntot_radial}
\end{figure*}

\section{Proxy map of CO (3--2)}
\label{sec:proxy3-2}

The CO spectral line energy distribution (SLED) presented in \citet{Esposito2024} reports line luminosities $L_{\rm CO(J{\to}J{-}1)}$ in solar units ($L_\odot$). To derive the brightness temperature line ratio $R_{32} \equiv L'_{\rm CO(3{-}2)} / L'_{\rm CO(2{-}1)}$, we convert between $L$ and $L'$ following \citet{Solomon2005}:
\begin{equation}
    L_{\rm CO(J{\to}J{-}1)} = 1.04 \times 10^{-3}\, \nu_{\rm rest}^3\, L'_{\rm CO(J{\to}J{-}1)},
    \label{eq:L_Lprime}
\end{equation}
where $L$ is in $L_\odot$, $\nu_{\rm rest}$ is the rest frequency in GHz, and $L'$ is in K\,km\,s$^{-1}$\,pc$^2$. Since $\nu_J = J \times \nu_{10}$ with $\nu_{10} = 115.271$\,GHz, the ratio $R_{32}$ becomes
\begin{equation}
    R_{32} = \frac{L'_{\rm CO(3{-}2)}}{L'_{\rm CO(2{-}1)}} = \frac{L_{\rm CO(3{-}2)}}{L_{\rm CO(2{-}1)}} \left(\frac{2}{3}\right)^3 = \frac{8}{27}\,\frac{L_{\rm CO(3{-}2)}}{L_{\rm CO(2{-}1)}}.
    \label{eq:R32}
\end{equation}

From Fig.\,C11 of \citet{Esposito2024}, the observed CO luminosities for NGC\,1275 are $L_{\rm CO(2{-}1)} \approx 1.5 \times 10^5\,L_\odot$ and $L_{\rm CO(3{-}2)} \approx 5 \times 10^5\,L_\odot$. Applying Eq.~\ref{eq:R32} yields $R_{32} \approx 1.25 \pm0.25$, consistent with the low-$J$ CO lines being close to thermalisation in the central region of NGC\,1275.
 
We note that these luminosities were compiled from observations with different beam sizes: the CO(2$-$1) from the IRAM 30\,m with a 10\farcs5 beam \citep{Lazareff1989}, and the CO(3$-$2) from the JCMT with a 21\arcsec\ beam \citep{Bridges1998}. This aperture mismatch introduces systematic uncertainties in the line ratios derived from the integrated CO SLED.

In addition, we computed the theoretical $R_{32}$ using \textsc{SpectralRadex} over a range of gas densities, kinetic temperatures, and column densities representative of the circumnuclear disc. The resulting distribution yields $R_{32} \approx 1.8^{+0.27}_{-0.35}$.
 
Combining the observational constraint from the CO SLED and the radiative transfer estimate, we construct two proxy CO(3--2) maps by scaling the observed CO(2--1) intensity with $R_{32} = 1.5$ and $R_{32} = 2.07$, respectively. Even under the most conservative correction, the enhanced HCO$^+$/CO ratio along the western edge of the disc persists. We stress that such a uniform scaling can only provide a first-order approximation of the CO(3--2) distribution, as spatial variations in excitation conditions across the disc are not captured. High-resolution CO(3--2) observations are needed to confirm this result.
\begin{figure*}
\centering
\includegraphics[width=8cm]{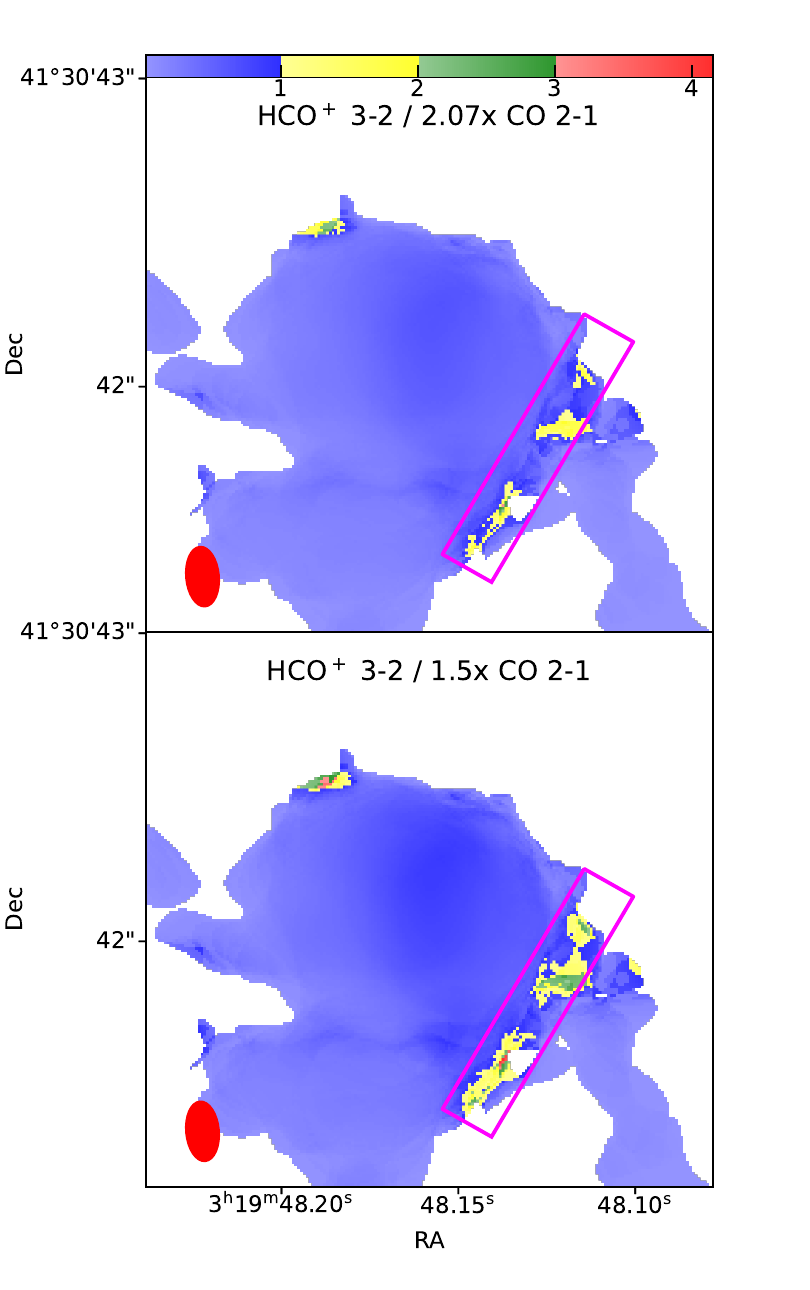}
\caption{Proxy map of $\mathrm{HCO}^{+}(3\text{--}2)/\mathrm{CO}(3\text{--}2)$ obtained by scaling the observed $\mathrm{CO}(2\text{--}1)$ intensity by a factor of 2.07 and 1.5, which are constrained from model and observation, respectively. }
\label{fig:hcop_co32_proxy}
\end{figure*}

\section{Kinematics of the molecular gas}
\label{app:kinematics}

We use the line profiles to test whether the three transitions trace the same gas in each region and to characterise the kinematics of the western HCO$^+$/CO enhancement. For each hexagon we fit a single Gaussian to the integrated CO(2--1), HCN(3--2), and HCO$^+$(3--2) spectra and record the centroid velocity $v_{\rm cen}$ and the full width at half maximum (FWHM). Regions in which a line falls below the $3\sigma$ detection threshold are excluded. We show line profiles for three regions that sample the centre, the disc, and the filament--disc interface (regions 1, 4, and 9 in Fig.~\ref{fig:galleryofNtotcheck}).

The velocity fields and position-velocity plots of the three transitions (Figs.~\ref{fig:velfies} and~\ref{fig:slices}) already show that they share the same large-scale kinematics across many positions in the disc. The profiles below test this at the scale of individual hexagons.

\begin{figure*}
\centering
\includegraphics[width=\textwidth]{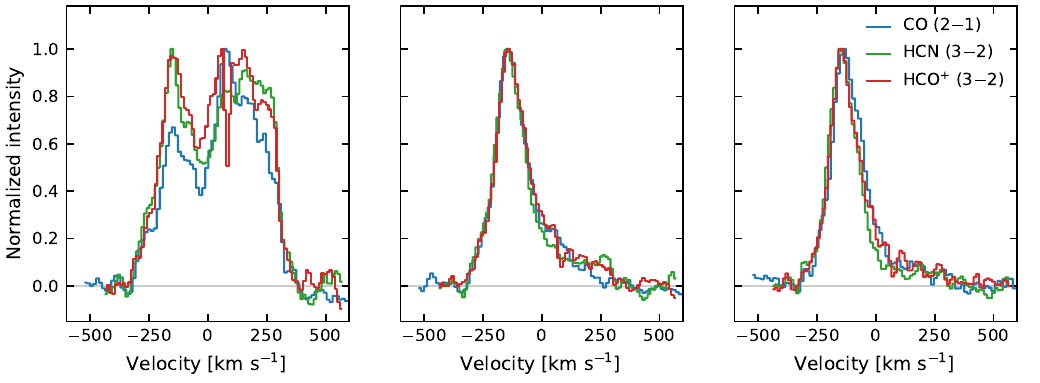}
\caption{Normalised CO(2--1), HCN(3--2), and HCO$^+$(3--2) line profiles for three regions that sample the centre, the disc, and the filament--disc interface (regions 1, 4, and 9 in Fig.~\ref{fig:galleryofNtotcheck}), shown left to right. Velocities are relative to the systemic velocity.}
\label{fig:spectra_kin}
\end{figure*}

Figure~\ref{fig:spectra_kin} shows the normalised profiles for these three regions, from the central region (left) to the filament--disc interface (right). In the central region all three lines are multi-peaked. A single 72\,pc beam there covers both the approaching and receding sides of the fast inner rotation, so the integrated spectrum splits into two peaks instead of resolving the velocity gradient. The profiles narrow to a single peak with increasing radius, and the three lines overlap in centroid and width. The multi-peaked centre is a caveat for the inferred parameters in the innermost hexagons, where a single-component model cannot capture the underlying velocity field.

\begin{figure*}
\centering
\includegraphics[width=0.85\textwidth]{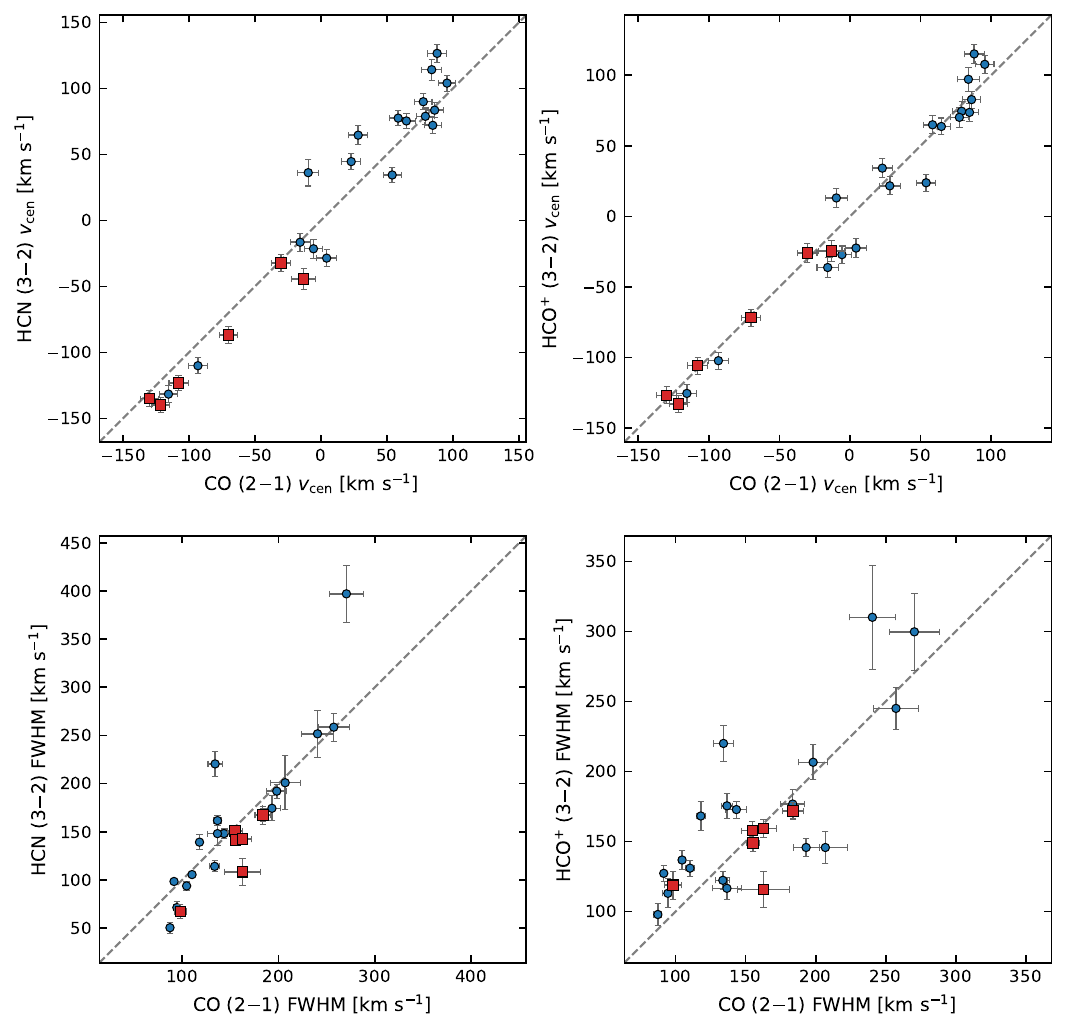}
\caption{Centroid velocity (top) and FWHM (bottom) of HCN(3--2) and HCO$^+$(3--2) against CO(2--1), from single-Gaussian fits to the integrated spectrum of each hexagon. The dashed line is the one-to-one relation. Red squares mark regions on the western side of the disc, and blue circles mark all other regions. Error bars show the $1\sigma$ fit uncertainties. Regions in which any line is undetected are excluded.}
\label{fig:scatter_kin}
\end{figure*}

Figure~\ref{fig:scatter_kin} compares the centroid velocities and line widths of the dense-gas tracers with those of CO across all detected regions. The centroids of HCN(3--2) and HCO$^+$(3--2) follow CO(2--1) along the one-to-one relation over the full velocity range, with a scatter consistent with the fitting uncertainties. The line widths agree to within about 30\,km\,s$^{-1}$ in most regions, with larger scatter than the centroids. The agreement in centroid velocity and line width indicates that the three transitions trace co-moving gas in each region and that no line is dominated by a kinematically distinct component along the line of sight. This hexagon-level agreement is consistent with the matching velocity fields of the three lines in Fig.~\ref{fig:velfies}. It is a necessary condition for treating the three integrated intensities as constraints on a single emitting region.

The western regions are marked in red in Fig.~\ref{fig:scatter_kin}. These are the hexagons on the western side, from the inner disc out to the filament--disc interface. Their centroid velocities are blueshifted, between about $-30$ and $-130$\,km\,s$^{-1}$, and lie on the same one-to-one relation as the rest of the disc. This is consistent with the blueshifted velocity of the western side of the rotating disc found by \citet{Oosterloo2024}, and the same regions show a local disturbance in the CO velocity field (Fig.~\ref{fig:velfies}). They tend toward larger single-Gaussian widths than the rest of the disc at comparable radius, seen as the red points at larger FWHM in Fig.~\ref{fig:scatter_kin}. The western profiles are not well described by a single Gaussian, so these widths overstate any intrinsic broadening and we read them as upper limits. Three points in the HCO$^+$(3--2) versus CO(2--1) panel lie well above the one-to-one relation. They correspond to regions where a single Gaussian poorly describes an asymmetric or double profile, and we exclude them from this comparison.

The local disturbance in the velocity field and the larger line widths both appear on scales at or below the beam size, so a distinct kinematic component at the filament--disc interface cannot be isolated within a single hexagon. The velocity dispersion at this resolution is dominated by unresolved rotation and beam smearing, so we do not use it as a diagnostic. We therefore treat the kinematics of the western enhancement as suggestive rather than conclusive given the current data.

\end{appendix}
\end{document}